\documentclass[prl, twocolumn, superscriptaddress, nofootinbib, nobibnotes]{revtex4-2}

\usepackage[normalem]{ulem}

\usepackage[english]{babel}
\usepackage{hyperref}

\usepackage[pdftex]{graphicx}
\usepackage{color}

\usepackage{amsmath}
\usepackage{amssymb}
\usepackage{mathtools}
\usepackage{braket}
\usepackage{stmaryrd}

\usepackage[sort&compress]{natbib}

\newcommand{\un}[1]{\ensuremath{\,\mathrm{#1}}}
\renewcommand{\v}[1]{\ensuremath{\boldsymbol{#1}}}

\newcommand{\fig}[1]{Figure~\ref{fig:#1}}

\newcommand{\lr}[1]{\ensuremath{\left( #1 \right)}}

\newcommand{\I}{\mathrm{i}}

\newcommand{\mc}{\mathcal}

\newcommand{\affil}{\affiliation{Instituto de Ciencias F\'isicas, Universidad Nacional Aut\'onoma de M\'exico, Cuernavaca, México}}
\newcommand{\afffil}{\affiliation{Facultad de Química, Universidad Nacional Aut\'onoma de M\'exico, Ciudad de México, México}}

\begin{document}

\title{Edge-state transport in twisted bilayer graphene}

\author{Jesús~Arturo~Sánchez-Sánchez}
\email{jasanchez@icf.unam.mx}
\affil

\author{Montserrat~Navarro-Espino}
\email{monsene10@gmail.com}
\afffil

\author{José~Eduardo~Barrios-Vargas}
\email{j.e.barrios@gmail.com}
\afffil

\author{Thomas~Stegmann}
\email{stegmann@icf.unam.mx}
\affil

\date{\today}

\begin{abstract}
We investigate the electronic structure and transport properties of twisted bilayer graphene (TBLG) at a twist angle of $\theta\approx 1.696\text{°}$. Using a combination of molecular dynamics and tight-binding calculations, we find two superlattice gaps in the energy spectrum of the bulk, which emerge close to the Fermi level from the atomic rearrangement of the carbon atoms leading to a corrugation of the graphene sheets. Nanoribbons made from 1.696°-TBLG show edge-localized states inside the superlattice gaps. Applying the Green's function method, we demonstrate that the edge states carry electronic current with conductance values close to the conductance quantum. The edge states can generate a non-local resistance, which is not due to one-way transport at the edges but due to the fact that these states are localized only at certain edges of the system, depending on how the nanoribbon has been cut from the bulk.
\end{abstract}

\maketitle

\section{Introduction}

\begin{figure*}[t]
   \centering
   \includegraphics[width=\linewidth]{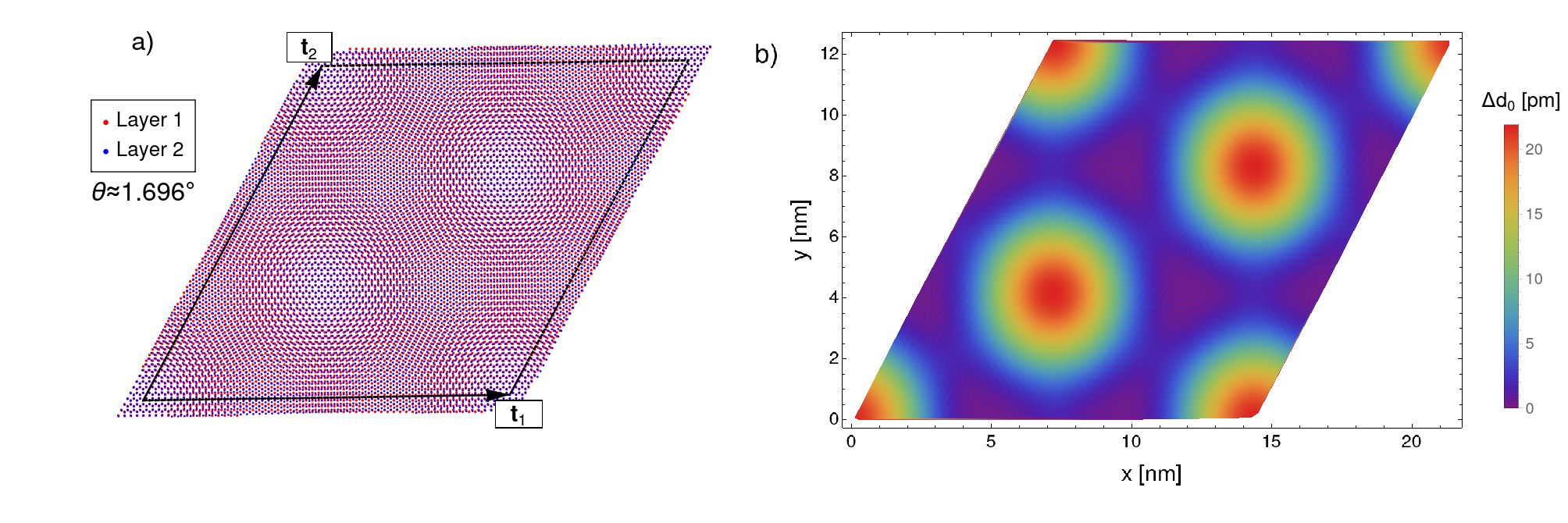}
\caption{(a) Supercell of the studied system, consisting of two stacked graphene layers with a twist angle of $\theta \approx 1.696\text{°}$. The resulting 1.696°-TBLG system shows a Moire pattern of alternating AA/AB stacking areas. (b) Change of the interlayer distance $\Delta d_0$ after performing a structural relaxation by the MD method. The two layers are no longer flat but corrugated.}
   \label{fig:sys}
\end{figure*}
 
Graphene paved the way for the emergence of 2D materials almost two decades ago \cite{1}, so it is not surprising that the graphene bilayer is doing the same for stacked structures made of 2D materials. Bilayer graphene (BLG) comes in two semi-metallic flavors, one with a parabolic low-energy dispersion \cite{2,3,4} in its most common and stable phase known as AB or Bernal stacking, and one with displaced Dirac cones in the less common and energetically unfavorable AA stacking phase \cite{AA1,AA2,AA3}. In general, BLG exhibits both excellent electrical and thermal conductivity at room temperature \cite{5,6}, mechanical stiffness, strength and flexibility \cite{7,8}, high transparency \cite{9} and allows the tuning of its electrical properties via external gating and doping \cite{10,11,12,13,naumisstrained}.

In recent years, introducing a twist angle between two stacked graphene layers has enabled an extraordinary array of physical phenomena not present in neither graphene nor BLG \cite{14}. The twist angle generates what is called a Moiré pattern (see \fig{sys}), an alternating distribution of stacking zones that alters the way the monolayers interact with each other \cite{15,16,TBLG2}. This interaction has been shown to lead to strongly correlated electronic behavior such as unconventional superconductivity and Mott-like insulating states \cite{17,18}. With twisted bilayer graphene (TBLG) at the forefront, the field of twistronics has emerged as a highly fertile research ground with promising technological applications \cite{TBLG3}.

One distinct feature of these TBLG systems is how the twist angle affects both the atomic arrangement and the electronic structure. It has been shown that an atomic reconstruction is energetically favorable for TBLG leading to a corrugation of the layers \cite{25,26,buckle1}, which is particularly steep for twist angles below 5°, whereas for angles above 15° the monolayers remain essentially flat. For certain small angles, the atomic reconstruction even leads to the creation of band gaps, named also superlattice gaps \cite{gaps1,gaps2}.  Graphene's dispersion remains linear for most twist angles \cite{19,20,TBLG1} and the van Hove singularities in its density of states are shifted towards energies which are accessible through gating \cite{15,21,22}. Furthermore, "magic" angles lead to the existence of flat bands where electrons are tightly packed in momentum space and correlation effects become important. Most of the research is focused on this magic angle TBLG, both theoretically \cite{macdonald,20,naumis2x2,MATBLG} and experimentally \cite{17,18,23}. However, interesting phenomena exist for many twist angles \cite{nosotros,gaps2,stuff1}.

Edge states in TBLG have attracted some attention. It has been demonstrated experimentally that the zigzag edge states of graphene can be tuned by stacking two zigzag nanoribbons with a twist \cite{edge2}. In magic angle TBLG edge states can give rise to ferromagnetism and the anomalous Hall effect \cite{edge1}. A variety of edge states were calculated in TBLG nanoribbons that reside amid other electronic states \cite{edge3} and it has been shown that the mechanical sliding of one layer in TBLG causes the migration of edge states in the energy spectrum \cite{slide}. Particularly important for this article are transport measurements by Ma et al. showing a non-local resistance in a TBLG device with a twist of 1.68°, which they attribute to the existence of counter-propagating edge states within the superlattice gaps \cite{24}.

In this paper, we aim to shed light on these edge states. We investigate TBLG at a twist angle of $\theta \approx 1.696\text{°}$ -- named in the following 1.696°-TBLG -- by using a combination of molecular dynamics (MD) and tight-binding (TB) methodologies. We find two superlattice gaps in the electronic structure of the bulk, which are due to the atomic reconstruction of the carbon atoms that lead to a corrugation of the graphene layers. In the case of nanoribbons made from 1.696°-TBLG, edge localized states reside inside these gaps. The nonequilibrium Green’s function (NEGF) method is applied to the MD+TB model to study the current flow in a 1.696°-TBLG device. We find that the edge states carry current with conductance values close to the conductance quantum and can lead to a nonlocal resistance. However, the edge states are bi-directional and the non-local resistance is due to the fact that the edge states are localized only at certain edges of the system, depending on how the nanoribbon is cut from the bulk. 

\section{System \& Methods}

\subsection{TBLG unit cell}
 
In order to construct the TBLG systems, we start with a graphene monolayer with lattice vectors $\mathbf{a}_1=a_0 ( \sqrt{3},0)$ and $\mathbf{a}_2=a_0(\frac{\sqrt{3}}{2},\frac{3}{2})$, where $a_0=0.142\un{nm}$ is the carbon-carbon distance. The atomic base consists of carbon atoms at positions $(0,0)$ and $\frac{1}{3}(\mathbf{a}_1 + \mathbf{a}_2)$. A Moiré pattern will emerge by taking two graphene layers, separated vertically by $d_{0}=0.335\un{nm}$, and rotating them by the twist angle 
\begin{equation}
    \label{theta}
\cos{\theta}=\frac{2p^2+2pq-q^2}{2(p^2+pq+q^2)}.
\end{equation}

The positive integers $p$ and $q$ will identify a given TBLG system and define the pair of superlattice vectors $\mathbf{t}_1=p\mathbf{a}_1+q\mathbf{a}_2$ and $\mathbf{t}_2=-q\mathbf{a}_1+(p+q)\mathbf{a}_2$. 

\fig{sys} (a) shows a TBLG system constructed by this method. The system has a twist angle of $\theta \approx 1.696\text{°}$, corresponding to the pair of integers $p=58$ and $q=1$. The new supercell, defined by the dashed parallelogram, contains a pattern of alternating stacking areas. The AA stacking is visible where carbon sites of one layer are directly above carbon sites of the other layer while the AB stacking can be seen where hexagons of one layer enclose a carbon site from the other layer. The continuous modulation of these stacking zones creates the Moiré pattern. Systems constructed from this particular TBLG supercell, containing 13,392 carbon atoms, will be the main focus of this article.
 
\subsection{Structural relaxation}

Two methods are typically used for the structural relaxation of the TBLG system. Density Functional Theory (DFT) calculations are accurate but computationally demanding and thus, limited to large twist angles where the supercell is rather small \cite{dft1}. Because MD calculations are able to handle hundreds of thousands of atoms and correctly reproduce the corrugation of the graphene layers \cite{25,buckle1,tb1}, we perform structural relaxation calculations on the TBLG supercell via the LAMMPS MD code \cite{lammps1,lammps2}. The interlayer REBO \cite{rebo1,rebo2} potential and the intralayer Kolmogorov-Crespi \cite{kolmogorov} potential are employed since they are tailored to graphitic and misalinged graphene layered systems, respectively. The new relaxed structure of the 1.696°-TBLG  supercell (see \fig{sys} (b)) will be utilized for the remainder of the article.
 
\subsection{Tight-Binding model}

\begin{figure*}[t]
  \centering
   \includegraphics[width=\linewidth]{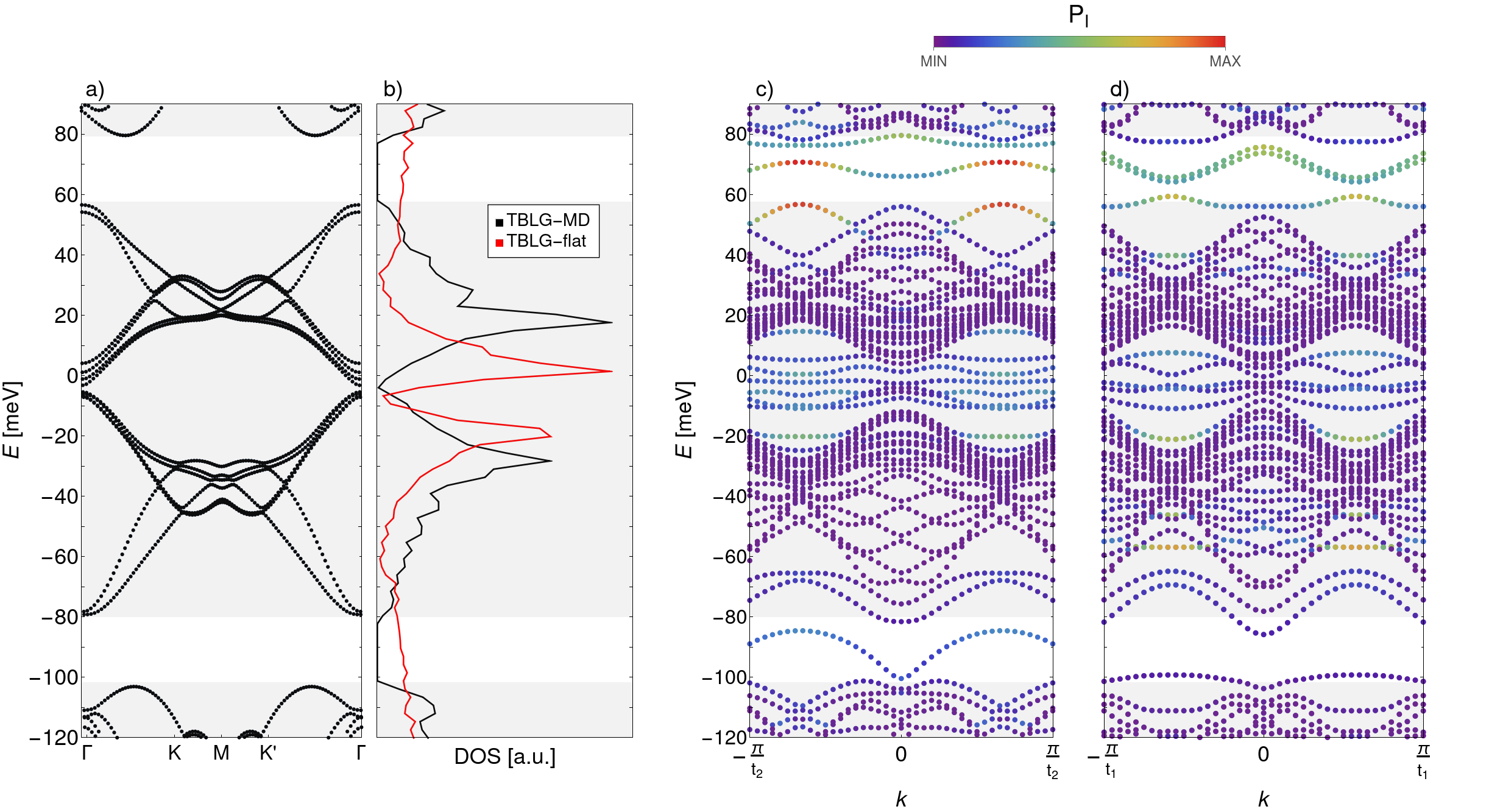}
   \caption{Electronic structure of 1.696°-TBLG systems. (a) Energy bands and (b) density of states (DOS) of the 1.696°-TBLG bulk system. Two energy gaps with a size of about 20 meV are found, which are absent in the flat system without MD optimization (red curve). (c,d) Energy bands of a 1.696°-TBLG nanoribbon with a width of 4 supercells in two periodic directions, $\mathbf{t}_2$ and $\mathbf{t}_1$, respectively. States appear inside the energy gaps. The color of the energy bands indicates their inverse participation ratio $P_\text{I}$ and demonstrates the localization of the states inside the gap.}
   \label{fig:Electronic}
\end{figure*}

Effective models and low energy approximations are able to reproduce key features of the electronic structure of TBLG bulk systems \cite{19,tb4,tb2,macdonald}, such as the renormalization of the Fermi velocity and flat bands. However, we are interested not only in bulk systems but also in nanoribbons and devices and choose for the TBLG system the TB Hamiltonian
\begin{equation}
    \label{H}
    H{}= \sum_{n,m} t_{nm} \ket{n} \bra{m} + \text{h.c.}
\end{equation}
where $\ket{n}$ indicates the atomic state localized on the carbon atom at $\v{r}_n$. The coupling term between sites $t_{nm}$ is determined by the Slater-Koster formula \cite{slater} 
\begin{equation}
    \label{t_nm}
    t_{nm}= \cos^{2} (\gamma) V_{pp\sigma}(r_{nm})+ [1-\cos^{2}(\gamma)]V_{pp\pi}(r_{nm})
\end{equation}
where $\cos\gamma=z_{nm}/r_{nm}$ is the direction cosine of $\v{r}_{nm}=\v{r}_m-\v{r}_n$ along the $z$ direction. The Slater-Koster parameters are defined as 
\begin{equation}
    \label{Slater-Koster}
    \begin{split}
    V_{pp\sigma}(r_{nm})&=V_{pp\sigma}^{0}\exp\left[q_{\sigma} \Bigl (1-\frac{r_{nm}}{d_0} \Bigr)\right],\\
    V_{pp\pi}(r_{nm})&=V_{pp\pi}^{0}\exp\left[q_{\pi} \Bigl(1-\frac{r_{nm}}{a_0} \Bigr) \right]
    \end{split}
\end{equation}
with the tight-binding parameters \cite{param} 
\begin{equation}
\label{TB_parameters}
    \begin{split}
    V_{pp\sigma}^{0}=0.48 \un{eV}, \quad V_{pp\pi}^{0}=-2.7 \un{eV},\\[1mm]
    \frac{q_{\pi}}{a_0}=\frac{q_{\sigma}}{d_0}=22.18 \un{nm}^{-1}. \quad
    \end{split}
\end{equation}

\subsection{Nonequilibrium Green's funcion method}

\begin{figure*}[t]
   \centering
   \includegraphics[width=\linewidth]{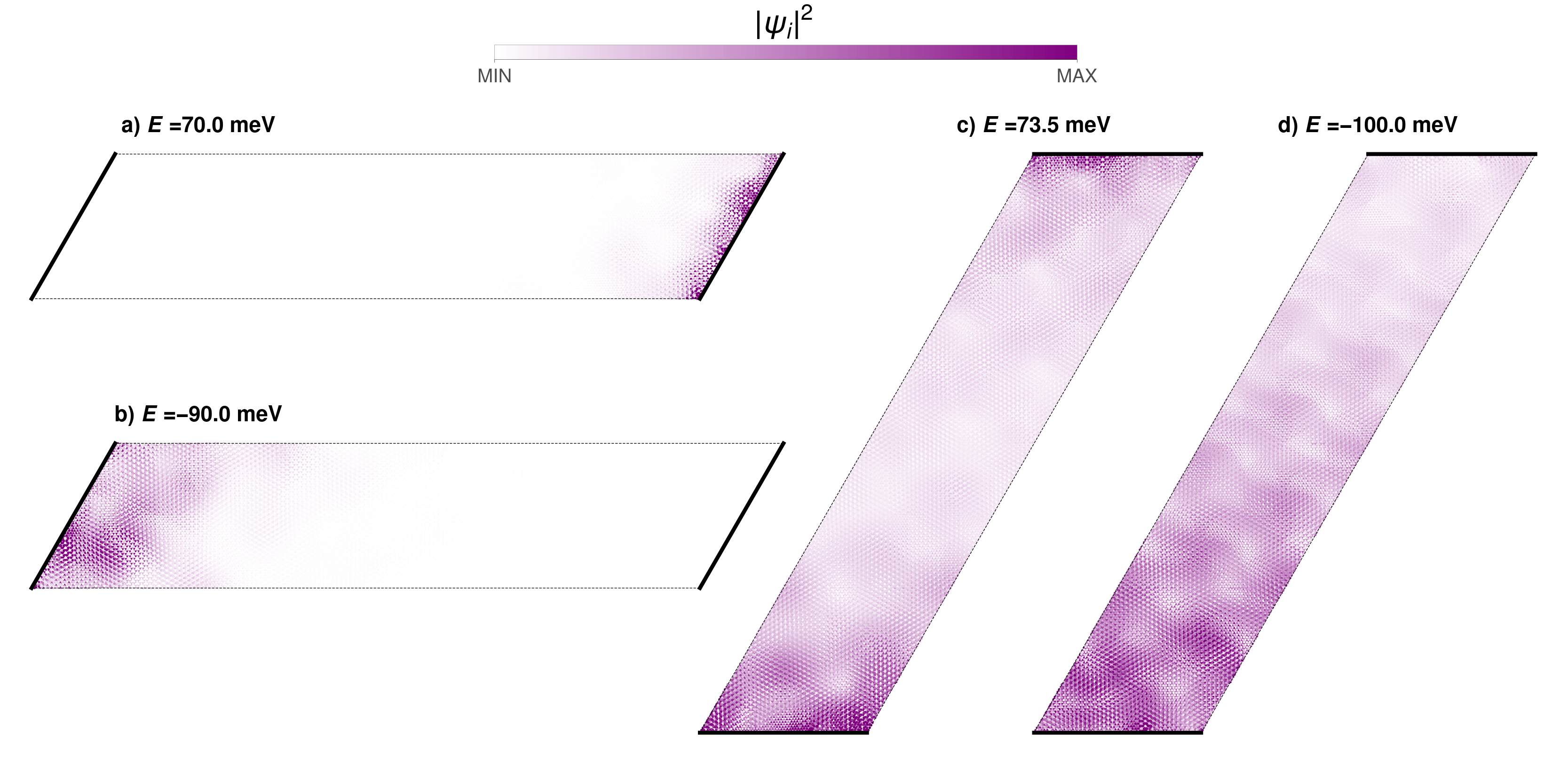}
   \caption{Eigenstates of the 1.696°-TBLG nanoribbons. 
   Solid thick lines indicate the edges of the nanoribbon, while the dashed thin lines show the periodic directions; in (a,b) the nanoribbons extend in the $\mathbf{t}_2$ direction whereas in (c,d) they are periodic in the $\mathbf{t}_1$ direction. (a-c) confirm that the states inside the superlattice gaps are localized at the edges of the nanoribbons. The state in (d) shows that at the edges of the bandgaps the state becomes delocalized.}
   \label{fig:Local}
 \end{figure*}

The NEGF method is applied to study the electrical transport in the TBLG devices and here, we briefly summarize the essential equations \cite{negf1,negf2,negf3}. The Green's function of the system is given by
\begin{equation}
    \label{GF}
    G=\Bigl(E-H-\sum_p\Sigma_p\Bigr)^{-1}
\end{equation}

\noindent where $E$ is the energy of the injected electrons and $H$ is the tight-binding Hamiltonian \eqref{H}. The contacts are modeled by the so-called wide-band model implying for these a constant energy-independent surface density of states. It is taken into account in the Green's function by means of the self-energies $ \Sigma_{p}= -\I V_{pp\pi}^{0} \sum_{n \in C_{p}} \ket{n}\bra{n}$, where the sum runs over the carbon atoms $C_p$ attached to the contact $p$.

The transmission between a pair of contacts, $i$ and $j$, is given by
\begin{equation}
    \label{Transmission}
	T_{ij}(E)= \mathrm{Tr}[\Gamma_{i}G\Gamma_{j}G^{\dagger}],
\end{equation}
where the inscattering function associated to contact $i$ is defined as $\Gamma_{i}= -2\mathrm{Im}(\Sigma_i)$. The transmission determines the conductance for electrons of energy $E$ between the selected pair of contacts, $\mathcal{G}_{ij}= (e^2/h) \, T_{ij}$. Furthermore, the local current flowing between the atoms $n$ and $m$ can be calculated with \cite{curr1,curr2}
\begin{equation}
    \label{I}
I_{nm}(E)= \frac{e}{h}\mathrm{Im}\bigl[t^{*}_{nm} (G\Gamma_{S}G^{\dagger})_{nm} \bigr].
\end{equation}

Finally, for the non-local resistance calculations we consider 4 contacts in the device, where electrons are injected by the S contact, collected in the D contact and the voltage between a pair of contacts A and B is calculated. The non-local resistance is calculated as follows
\begin{align}
    \label{Rnl}
R_{\text{NL}}&=\frac{V_{AB}}{I_{SD}}=\frac{h}{e^2} \frac{\sum_j T_{Sj} \lr{\mc{M}_{jA}-\mc{M}_{j B}}}{T_{SD} + \sum_{ij} T_{Di} \mc{M}_{ij} T_{jS}}
\end{align}
where
\begin{equation}
    \label{Mmat}
    \mc{M}^{-1}=
   \begin{cases}
   -T_{ij} \quad &\text{if} \quad i\neq j,\\
   \sum_{k\neq i} T_{ik} \quad &\text{if} \quad i=j.
   \end{cases}
\end{equation}
For more details on the derivation of \eqref{Rnl} and \eqref{Mmat} in the context of the NEGF method, see our previous work \cite{nosotros}.

\section{Band structure calculations}

\begin{figure*}[t]
   \centering
   \includegraphics[width=\linewidth]{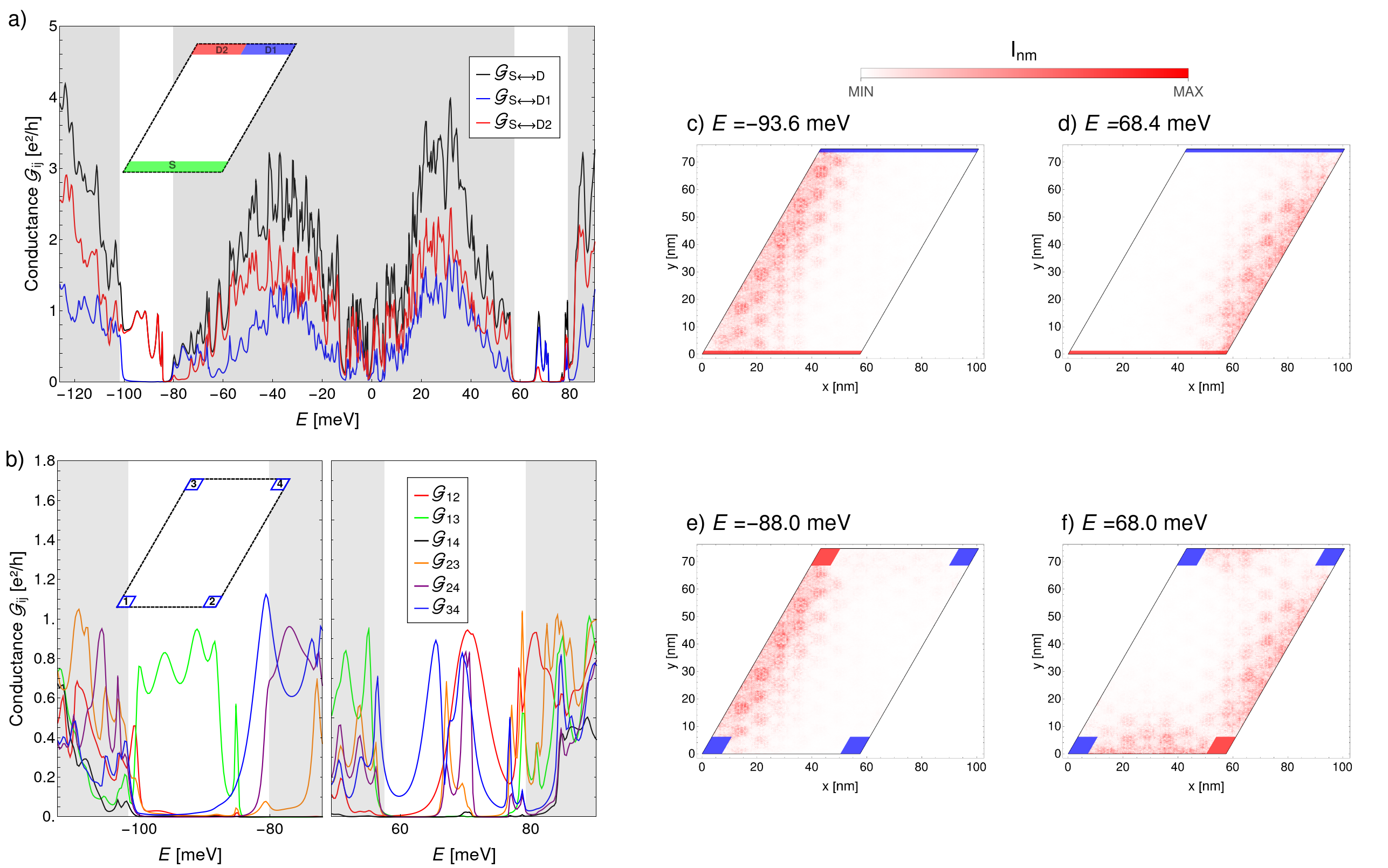}
   \caption{Electronic transport in a 1.696°-TBLG device with a size of $4\text{x}6$ supercells. (a) Conductance as a function of the electron energy for the device with 3 contacts as sketched in the inset. The colors of the conductance curves match the colors of the drain contacts. The black curve gives the conductance in the case of only a single drain contact $D=D_1\cup D_2$. (b) conductance in a device with 4 contacts at the corners that allow to identify transport along all edges. Both panels confirm that transport takes place inside the superlattice gaps with conductance values close to the conductance quantum. (c-f) The current density for electrons, injected by the red-marked source contacts, confirms the edge dominated transport inside the superlattice gaps.}
   \label{fig:Transport}
\end{figure*}

We perform the MD+TB calculation to obtain the electronic structure of the 1.696°-TBLG bulk system, shown in \fig{Electronic} (a,b). Two band gaps with a size of about 20 meV, highlighted by the white shaded regions, are found. The DOS, which is calculated over the full Brillouin zone, confirms these gaps. We have included also the DOS of a flat system, where MD calculations were not performed and no energy gaps are found. Hence, the energy gaps in \fig{Electronic} (a) and (b) originate from the atomic reconstruction of the system at such a small twist angle and we will refer to them as superlattice gaps for the remainder of the article. Note that two van Hove singularities are present at low energies, a fingerprint of the TBLG systems. Also note that given the TBLG supercell used for the calculations, see \fig{sys}, the Brillouin Zone is folded and the Dirac crossing is moved from the $K$ and $K'$ points to $\Gamma$.

With the bulk energy bands as a reference point, we turn our attention to the energy landscape of nanoribbons. \fig{Electronic} (c) and (d) show the energy bands of two 1.696°-TBLG nanoribbons, both with a width of 4 supercells and a total of $N=54,768$ atoms each. Nanoribbon (c) is periodic in the direction of vector $\mathbf{t}_2$ while nanoribbon (d) is periodic in the direction of $\mathbf{t}_1$. We find for these nanoribbons states inside the superlattice gaps. The color of the energy bands gives the inverse participation ratio $P_I=\sum_{i=1}^N |\psi_{i}|^4$ at each ($E,k$) value \cite{ipr} and unveils the localization of the states in the superlattice gaps. Note that these states exist for both directions of the nanoribbon though the number of the states change.

The spatial distribution of the states inside the superlattice gaps of the 1.696°-TBLG nanoribbons are shown in \fig{Local}. Solid thick lines indicate the edges of the nanoribbons, while dashed thin lines mark their periodic directions; panels (a,b) correspond to the nanoribbon periodic in the $\mathbf{t}_2$ direction, while (c,d) are for the nanoribbon with periodicity in the $\mathbf{t}_1$ direction. The states in (a-c) are localized predominantly at the edges of the nanoribbon. States in (a,b) are limited exclusively to one edge, while the state in (c) is present at both edges. The state in (d) is included as a contrast showing a largely delocalized state. Note that in general the observed localization (and delocalization) of the eigenstates is in line with the values of the inverse participation ratio in \fig{Electronic} (c,d), since the higher the value of $P_I$ the smaller the spatial extension of the state and vice-versa.

So far, our MD+TB calculations performed on 1.696°-TBLG systems confirm the emergence of superlattice gaps due to the corrugation of the graphene layers. Furthermore, inside these superlattice gaps we find localized states at the edges of the nanoribbons. These results are in agreement with the findings in \cite{24}. Next, we explore the transport properties of these edge localized states.

\section{Electronic transport}

\begin{figure*}[t]
   \centering
\includegraphics[width=\linewidth]{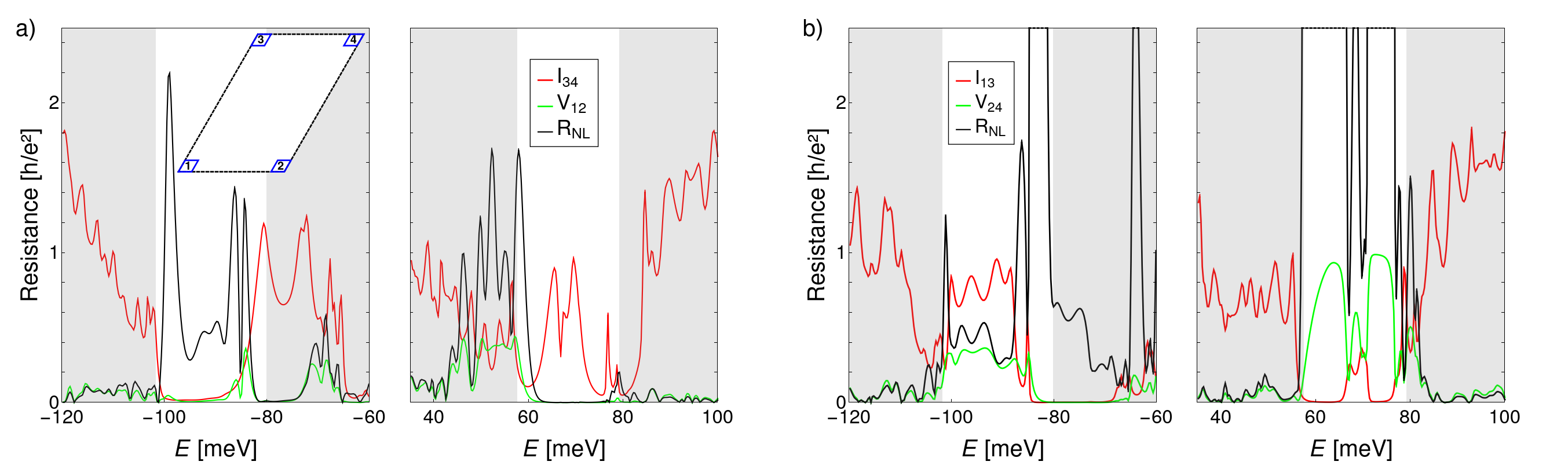}
   \caption{Non-local resistance calculations $R_{\text{NL}}$ for the 1.696°-TBLG device with dimensions of $4\text{x}6$ supercells. A finite non-local resistance is observed within the superlattice gaps for two different contact wirings in (a,b). $R_{\text{NL}}$ in the upper gap of panel (a) extends outside the gap, which can be understood by the fact that localized states exist also outside the gap (see \fig{Electronic}). In (b) $R_{\text{NL}}$ shows enormous values which can be explained by the fact that a voltage drop is detected (green curve) but no current is flowing though the device (red curve).}
   \label{fig:Res}
 \end{figure*}

We now turn our attention to the transport properties of the edge localized states inside the superlattice gaps. \fig{Transport} (a) and (b) show the conductance as a function of the electron energy for a 1.696°-TBLG device consisting of $4\text{x}6$ supercells. The contact positions are sketched in the insets. As in \fig{Electronic}, the background shading highlights the superlattice gaps and helps to identify transport in those. In panel (a) we observe transport inside the superlattice gaps with conductance values close to the conductance quantum $e^2/h$. We can also observe that the current is collected mainly, either by the left or right part of the drain contact, $D_2$ or $D_1$, depending on the superlattice gap. In panel (b) the contacts are placed at the corners of the nanoribbon, which allows to identify transport at all edges. Again, we observe that the current inside the superlattice gaps is carried predominantly along the edges, while the transport across the device, $\mathcal{G}_{14}$ and $\mathcal{G}_{23}$, is heavily suppressed. Panels (c-f) show the local current flow at certain electron energies and confirm the localization of the current on the edges of the system. This observation is in-line with the localization of the eigenstates shown in \fig{Local}. Importantly, we have verified in all our transport calculations that the edge states carry the current equally in both direction, $\mathcal{G}_{i\rightarrow{}j} = \mathcal{G}_{j\rightarrow{}i}\equiv \mathcal{G}_{ij}$, which is in agreement with the fact that time-reversal symmetry is preserved. 

Next, we analyze in \fig{Res} the non-local resistance of the 1.696°-TBLG device, consisting of $4\text{x}6$ supercells with 4 contacts. For that, the current flow is calculated between two contacts on one edge of the system, while the voltage drop is determined between two contacts on the opposite edge. The nonlocal resistance is defined as the quotient of these two quantities. In \fig{Res}, we show the voltage, current and the non-local resistance for two different contact wirings; in panel (a) the measurements are along the $\mathbf{t}_1$ direction while in (b) they are in the $\mathbf{t}_2$ direction. Both panels show within the superlattice gap a non-local resistance but the value of the resistance depends strongly on the wiring of the contacts. While in panel (a) values in the range of the resistance quantum are observed, the values in panel (b) can be extremely large. These large values can be explained by the fact that a voltage drop is detected between the contacts (green curves) but the current flowing though the device is almost zero (red curves). In the upper superlattice gap of panel (a), the non-local resistance extends partially to the region outside the gap, which can be understood by means of \fig{Electronic} showing that localized edge states appear also outside the gap. We find in \fig{Res} clear evidence for a nonlocal resistance inside the superlattice gap. However, we point out again that the transport along the edge states is bidirectional and the non-local resistance is due to the fact that the edge states are localized only on certain edges of the device, which will be discussed in the following.

\section{Nanoribbon edge termination}

 \begin{figure*}[t]
   \centering
   \includegraphics[width=\linewidth]{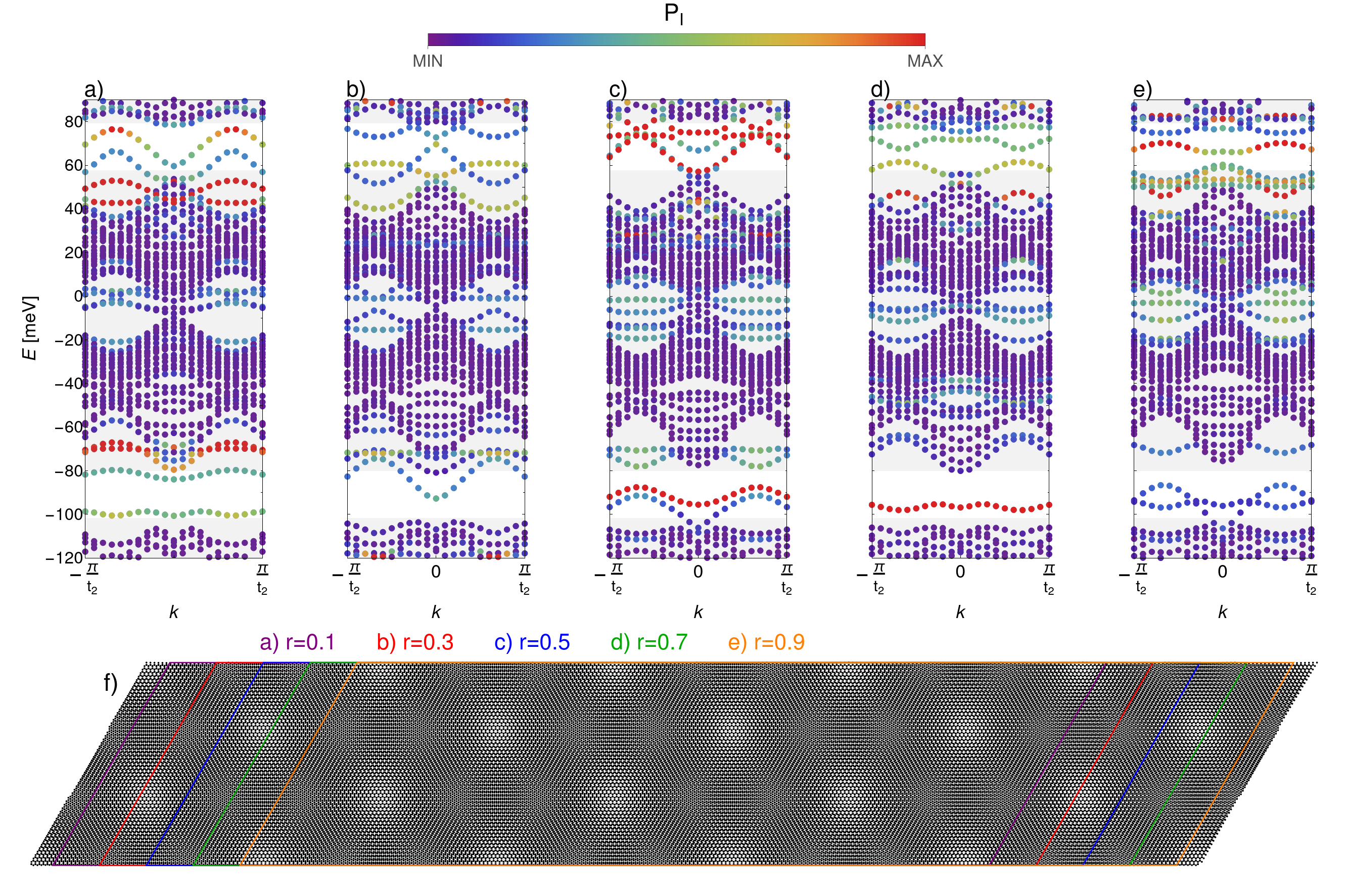}
   \caption{Band structure of nanoribbons with different edge termination. (a-e) Band structures of 1.696°-TBLG nanoribbons with a width of 4 unit cells and different edges. The nanoribbons are constructed by sliding a "cookie cutter" by $r\mathbf{t}_1$, $r= \{0.1,0.3,0.5,0.7,0.9\}$, over a 5 unit cell wide nanoribbon. As the edge varies, localized states migrate in and out of the superlattice gaps while unlocalized states essentially remain unchanged.}
   \label{fig:slide}
 \end{figure*}

We are left to understand the localization of the edge states at only certain edges of the system. In \fig{Electronic} (c,d), we analyzed the band structure of 1.696°-TBLG nanoribbons with a width of 4 unit cells and periodicity in the directions $\mathbf{t}_1$ and $\mathbf{t}_2$, respectively. These two nanoribbons showed differences in the number and position of edge localized states. Here, we examine the effect of the edge termination of the nanoribbon on its electronic structure. \fig{slide} shows how we construct the different versions of the nanoribbon by sliding a 4 unit cell wide "cookie cutter" over a nanoribbon with a width of 5 unit cells and periodicity in the $\mathbf{t}_2$ direction.  The sliding of the "cookie cutter" is given by $r\mathbf{t}_1$, with $r=\{ 0.1, 0.3, 0.5, 0.7, 0.9\}$. In this way, the 1.696°-TBLG  nanoribbon with a width of 4 unit cells is conserved but the edge termination changes. The case of $r=0$ corresponds for the already studied nanoribbon of \fig{Electronic} (c). Band structures for each nanoribbon are shown in \fig{slide} (a-e). The band structures show that the unlocalized (purple) bands remain essentially unchanged as we change the edges. In contrast, localized states enter and exit the superlattice gaps as the edge termination is modified. This migration of localized states in and out of the superlattice gaps is very akin to the results found in \cite{slide}. Regardless of the nanoribbon termination, \fig{slide} shows that one can expect a varying number of localized states inside the superlattice gaps of 1.696°-TBLG nanoribbons. When analyzing by eye the supercell shown in \fig{sys} one could argue that it may be identical to a supercell rotated by 180°. While this is true for the approximate position of the AA and AB stacking zones, this symmetry does not hold on the microscopic level of individual atoms and therefore, permits edge states only at certain edges. 

\section{Conclusions}

In this article, we have used a combination of TB, MD, and NEGF methodologies to investigate the electronic structure and transport properties of TBLG systems with a twist angle of $\theta \approx 1.696\text{°}$. Our calculations show that two superlattice gaps emerge in the energy bands of the bulk, see \fig{Electronic} (a), as a result of the atomic reconstruction that the system undergoes at small twist angles in order to minimize its energy. Using the relaxed supercell to build nanoribbons, \fig{Electronic} (b,c) show that states appear inside the superlattice gaps of the bulk. These states are localized predominately at the edges of the nanoribbons; see \fig{Local} (a-c). The edge localized states are dispersive and thus, able to carry electronic current as demonstrated by conductance and current density calculations for a device consisting of $4\text{x}6$ supercells (see \fig{Transport}). A non-local resistance is found within the superlattice gaps, though its magnitude depends on the contact wiring, see \fig{Res}. This non-local resistance, observed experimentally by Ma et al. \cite{24}, is not due to one-way transport at the edges but due to the fact that the edge states are localized only at certain edges of the system. Finally, in \fig{slide}, we demonstrate how different edge terminations of the nanoribbon lead to the appearance and disappearance of edge localized states across the superlattice gaps. Therefore, the edge states in $1.696\text{°}$-TBLG are not topological (in the sense of a bulk-boundary correspondence) but rather have the character of current-carrying states that are induced by edge disorder.

\section{Acknowledgments}
We thank L.E.F. Foa-Torres for useful discussions and comments. We also thank L. Brendel for his help with the LAMMPS code. JAS-S gratefully acknowledges a CONAHCYT graduate scholarship. MN-E acknowledges support from Subprograma 127-FQ, UNAM. We gratefully acknowledge funding from CONAHCYT Frontera 428214, UNAM-PAPIIT IN103922, UNAM-PAPIIT IA106021 and PAIP-FQ(UNAM) 5000-9173.

\vspace*{-5mm}
\enlargethispage{10mm}


\begin{thebibliography}{62}%
\makeatletter
\providecommand \@ifxundefined [1]{%
 \@ifx{#1\undefined}
}%
\providecommand \@ifnum [1]{%
 \ifnum #1\expandafter \@firstoftwo
 \else \expandafter \@secondoftwo
 \fi
}%
\providecommand \@ifx [1]{%
 \ifx #1\expandafter \@firstoftwo
 \else \expandafter \@secondoftwo
 \fi
}%
\providecommand \natexlab [1]{#1}%
\providecommand \enquote  [1]{``#1''}%
\providecommand \bibnamefont  [1]{#1}%
\providecommand \bibfnamefont [1]{#1}%
\providecommand \citenamefont [1]{#1}%
\providecommand \href@noop [0]{\@secondoftwo}%
\providecommand \href [0]{\begingroup \@sanitize@url \@href}%
\providecommand \@href[1]{\@@startlink{#1}\@@href}%
\providecommand \@@href[1]{\endgroup#1\@@endlink}%
\providecommand \@sanitize@url [0]{\catcode `\\12\catcode `\$12\catcode
  `\&12\catcode `\#12\catcode `\^12\catcode `\_12\catcode `\%12\relax}%
\providecommand \@@startlink[1]{}%
\providecommand \@@endlink[0]{}%
\providecommand \url  [0]{\begingroup\@sanitize@url \@url }%
\providecommand \@url [1]{\endgroup\@href {#1}{\urlprefix }}%
\providecommand \urlprefix  [0]{URL }%
\providecommand \Eprint [0]{\href }%
\providecommand \doibase [0]{https://doi.org/}%
\providecommand \selectlanguage [0]{\@gobble}%
\providecommand \bibinfo  [0]{\@secondoftwo}%
\providecommand \bibfield  [0]{\@secondoftwo}%
\providecommand \translation [1]{[#1]}%
\providecommand \BibitemOpen [0]{}%
\providecommand \bibitemStop [0]{}%
\providecommand \bibitemNoStop [0]{.\EOS\space}%
\providecommand \EOS [0]{\spacefactor3000\relax}%
\providecommand \BibitemShut  [1]{\csname bibitem#1\endcsname}%
\let\auto@bib@innerbib\@empty
\bibitem [{\citenamefont {Novoselov}\ \emph {et~al.}(2004)\citenamefont
  {Novoselov}, \citenamefont {Geim}, \citenamefont {Morozov}, \citenamefont
  {Jiang}, \citenamefont {Zhang}, \citenamefont {Dubonos}, \citenamefont
  {Grigorieva},\ and\ \citenamefont {Firsov}}]{1}%
  \BibitemOpen
  \bibfield  {author} {\bibinfo {author} {\bibfnamefont {K.~S.}\ \bibnamefont
  {Novoselov}}, \bibinfo {author} {\bibfnamefont {A.~K.}\ \bibnamefont {Geim}},
  \bibinfo {author} {\bibfnamefont {S.~V.}\ \bibnamefont {Morozov}}, \bibinfo
  {author} {\bibfnamefont {D.-e.}\ \bibnamefont {Jiang}}, \bibinfo {author}
  {\bibfnamefont {Y.}~\bibnamefont {Zhang}}, \bibinfo {author} {\bibfnamefont
  {S.~V.}\ \bibnamefont {Dubonos}}, \bibinfo {author} {\bibfnamefont {I.~V.}\
  \bibnamefont {Grigorieva}},\ and\ \bibinfo {author} {\bibfnamefont {A.~A.}\
  \bibnamefont {Firsov}},\ }\bibfield  {title} {\bibinfo {title} {Electric
  field effect in atomically thin carbon films},\ }\href
  {https://doi.org/10.1126/science.1102896} {\bibfield  {journal} {\bibinfo
  {journal} {science}\ }\textbf {\bibinfo {volume} {306}},\ \bibinfo {pages}
  {666} (\bibinfo {year} {2004})}\BibitemShut {NoStop}%
\bibitem [{\citenamefont {McCann}\ and\ \citenamefont {Koshino}(2013)}]{2}%
  \BibitemOpen
  \bibfield  {author} {\bibinfo {author} {\bibfnamefont {E.}~\bibnamefont
  {McCann}}\ and\ \bibinfo {author} {\bibfnamefont {M.}~\bibnamefont
  {Koshino}},\ }\bibfield  {title} {\bibinfo {title} {The electronic properties
  of bilayer graphene},\ }\href {https://doi.org/10.1088/0034-4885/76/5/056503}
  {\bibfield  {journal} {\bibinfo  {journal} {Reports on Progress in physics}\
  }\textbf {\bibinfo {volume} {76}},\ \bibinfo {pages} {056503} (\bibinfo
  {year} {2013})}\BibitemShut {NoStop}%
\bibitem [{\citenamefont {Novoselov}\ \emph {et~al.}(2006)\citenamefont
  {Novoselov}, \citenamefont {McCann}, \citenamefont {Morozov}, \citenamefont
  {Fal’ko}, \citenamefont {Katsnelson}, \citenamefont {Zeitler},
  \citenamefont {Jiang}, \citenamefont {Schedin},\ and\ \citenamefont
  {Geim}}]{3}%
  \BibitemOpen
  \bibfield  {author} {\bibinfo {author} {\bibfnamefont {K.~S.}\ \bibnamefont
  {Novoselov}}, \bibinfo {author} {\bibfnamefont {E.}~\bibnamefont {McCann}},
  \bibinfo {author} {\bibfnamefont {S.}~\bibnamefont {Morozov}}, \bibinfo
  {author} {\bibfnamefont {V.~I.}\ \bibnamefont {Fal’ko}}, \bibinfo {author}
  {\bibfnamefont {M.}~\bibnamefont {Katsnelson}}, \bibinfo {author}
  {\bibfnamefont {U.}~\bibnamefont {Zeitler}}, \bibinfo {author} {\bibfnamefont
  {D.}~\bibnamefont {Jiang}}, \bibinfo {author} {\bibfnamefont
  {F.}~\bibnamefont {Schedin}},\ and\ \bibinfo {author} {\bibfnamefont
  {A.}~\bibnamefont {Geim}},\ }\bibfield  {title} {\bibinfo {title}
  {Unconventional quantum hall effect and berry’s phase of 2$\pi$ in bilayer
  graphene},\ }\href {https://doi.org/https://doi.org/10.1038/nphys245}
  {\bibfield  {journal} {\bibinfo  {journal} {Nature physics}\ }\textbf
  {\bibinfo {volume} {2}},\ \bibinfo {pages} {177} (\bibinfo {year}
  {2006})}\BibitemShut {NoStop}%
\bibitem [{\citenamefont {Katsnelson}\ \emph {et~al.}(2006)\citenamefont
  {Katsnelson}, \citenamefont {Novoselov},\ and\ \citenamefont {Geim}}]{4}%
  \BibitemOpen
  \bibfield  {author} {\bibinfo {author} {\bibfnamefont {M.}~\bibnamefont
  {Katsnelson}}, \bibinfo {author} {\bibfnamefont {K.}~\bibnamefont
  {Novoselov}},\ and\ \bibinfo {author} {\bibfnamefont {A.}~\bibnamefont
  {Geim}},\ }\bibfield  {title} {\bibinfo {title} {Chiral tunnelling and the
  klein paradox in graphene},\ }\href
  {https://doi.org/https://doi.org/10.1038/nphys384} {\bibfield  {journal}
  {\bibinfo  {journal} {Nature physics}\ }\textbf {\bibinfo {volume} {2}},\
  \bibinfo {pages} {620} (\bibinfo {year} {2006})}\BibitemShut {NoStop}%
\bibitem [{\citenamefont {Berashevich}\ and\ \citenamefont
  {Chakraborty}(2011)}]{AA1}%
  \BibitemOpen
  \bibfield  {author} {\bibinfo {author} {\bibfnamefont {J.}~\bibnamefont
  {Berashevich}}\ and\ \bibinfo {author} {\bibfnamefont {T.}~\bibnamefont
  {Chakraborty}},\ }\bibfield  {title} {\bibinfo {title} {Interlayer repulsion
  and decoupling effects in stacked turbostratic graphene flakes},\ }\href
  {https://doi.org/https://doi.org/10.1103/PhysRevB.84.033403} {\bibfield
  {journal} {\bibinfo  {journal} {Physical Review B}\ }\textbf {\bibinfo
  {volume} {84}},\ \bibinfo {pages} {033403} (\bibinfo {year}
  {2011})}\BibitemShut {NoStop}%
\bibitem [{\citenamefont {Rakhmanov}\ \emph {et~al.}(2012)\citenamefont
  {Rakhmanov}, \citenamefont {Rozhkov}, \citenamefont {Sboychakov},\ and\
  \citenamefont {Nori}}]{AA2}%
  \BibitemOpen
  \bibfield  {author} {\bibinfo {author} {\bibfnamefont {A.}~\bibnamefont
  {Rakhmanov}}, \bibinfo {author} {\bibfnamefont {A.}~\bibnamefont {Rozhkov}},
  \bibinfo {author} {\bibfnamefont {A.}~\bibnamefont {Sboychakov}},\ and\
  \bibinfo {author} {\bibfnamefont {F.}~\bibnamefont {Nori}},\ }\bibfield
  {title} {\bibinfo {title} {Instabilities of the a a-stacked graphene
  bilayer},\ }\href
  {https://doi.org/https://doi.org/10.1103/PhysRevLett.109.206801} {\bibfield
  {journal} {\bibinfo  {journal} {Physical Review Letters}\ }\textbf {\bibinfo
  {volume} {109}},\ \bibinfo {pages} {206801} (\bibinfo {year}
  {2012})}\BibitemShut {NoStop}%
\bibitem [{\citenamefont {Roy}\ \emph {et~al.}(1998)\citenamefont {Roy},
  \citenamefont {Kallinger},\ and\ \citenamefont {Sattler}}]{AA3}%
  \BibitemOpen
  \bibfield  {author} {\bibinfo {author} {\bibfnamefont {H.-V.}\ \bibnamefont
  {Roy}}, \bibinfo {author} {\bibfnamefont {C.}~\bibnamefont {Kallinger}},\
  and\ \bibinfo {author} {\bibfnamefont {K.}~\bibnamefont {Sattler}},\
  }\bibfield  {title} {\bibinfo {title} {Study of single and multiple foldings
  of graphitic sheets},\ }\href
  {https://doi.org/https://doi.org/10.1016/S0039-6028(97)01032-7} {\bibfield
  {journal} {\bibinfo  {journal} {Surface science}\ }\textbf {\bibinfo {volume}
  {407}},\ \bibinfo {pages} {1} (\bibinfo {year} {1998})}\BibitemShut {NoStop}%
\bibitem [{\citenamefont {Dean}\ \emph {et~al.}(2010)\citenamefont {Dean},
  \citenamefont {Young}, \citenamefont {Meric}, \citenamefont {Lee},
  \citenamefont {Wang}, \citenamefont {Sorgenfrei}, \citenamefont {Watanabe},
  \citenamefont {Taniguchi}, \citenamefont {Kim}, \citenamefont {Shepard} \emph
  {et~al.}}]{5}%
  \BibitemOpen
  \bibfield  {author} {\bibinfo {author} {\bibfnamefont {C.~R.}\ \bibnamefont
  {Dean}}, \bibinfo {author} {\bibfnamefont {A.~F.}\ \bibnamefont {Young}},
  \bibinfo {author} {\bibfnamefont {I.}~\bibnamefont {Meric}}, \bibinfo
  {author} {\bibfnamefont {C.}~\bibnamefont {Lee}}, \bibinfo {author}
  {\bibfnamefont {L.}~\bibnamefont {Wang}}, \bibinfo {author} {\bibfnamefont
  {S.}~\bibnamefont {Sorgenfrei}}, \bibinfo {author} {\bibfnamefont
  {K.}~\bibnamefont {Watanabe}}, \bibinfo {author} {\bibfnamefont
  {T.}~\bibnamefont {Taniguchi}}, \bibinfo {author} {\bibfnamefont
  {P.}~\bibnamefont {Kim}}, \bibinfo {author} {\bibfnamefont {K.~L.}\
  \bibnamefont {Shepard}}, \emph {et~al.},\ }\bibfield  {title} {\bibinfo
  {title} {Boron nitride substrates for high-quality graphene electronics},\
  }\href {https://doi.org/10.1038/nnano.2010.172} {\bibfield  {journal}
  {\bibinfo  {journal} {Nature nanotechnology}\ }\textbf {\bibinfo {volume}
  {5}},\ \bibinfo {pages} {722} (\bibinfo {year} {2010})}\BibitemShut {NoStop}%
\bibitem [{\citenamefont {Ghosh}\ \emph {et~al.}(2010)\citenamefont {Ghosh},
  \citenamefont {Bao}, \citenamefont {Nika}, \citenamefont {Subrina},
  \citenamefont {Pokatilov}, \citenamefont {Lau},\ and\ \citenamefont
  {Balandin}}]{6}%
  \BibitemOpen
  \bibfield  {author} {\bibinfo {author} {\bibfnamefont {S.}~\bibnamefont
  {Ghosh}}, \bibinfo {author} {\bibfnamefont {W.}~\bibnamefont {Bao}}, \bibinfo
  {author} {\bibfnamefont {D.~L.}\ \bibnamefont {Nika}}, \bibinfo {author}
  {\bibfnamefont {S.}~\bibnamefont {Subrina}}, \bibinfo {author} {\bibfnamefont
  {E.~P.}\ \bibnamefont {Pokatilov}}, \bibinfo {author} {\bibfnamefont {C.~N.}\
  \bibnamefont {Lau}},\ and\ \bibinfo {author} {\bibfnamefont {A.~A.}\
  \bibnamefont {Balandin}},\ }\bibfield  {title} {\bibinfo {title} {Dimensional
  crossover of thermal transport in few-layer graphene},\ }\href
  {https://doi.org/https://doi.org/10.1038/nmat2753} {\bibfield  {journal}
  {\bibinfo  {journal} {Nature materials}\ }\textbf {\bibinfo {volume} {9}},\
  \bibinfo {pages} {555} (\bibinfo {year} {2010})}\BibitemShut {NoStop}%
\bibitem [{\citenamefont {Neek-Amal}\ and\ \citenamefont {Peeters}(2010)}]{7}%
  \BibitemOpen
  \bibfield  {author} {\bibinfo {author} {\bibfnamefont {M.}~\bibnamefont
  {Neek-Amal}}\ and\ \bibinfo {author} {\bibfnamefont {F.}~\bibnamefont
  {Peeters}},\ }\bibfield  {title} {\bibinfo {title} {Nanoindentation of a
  circular sheet of bilayer graphene},\ }\href
  {https://doi.org/https://doi.org/10.1103/PhysRevB.81.235421} {\bibfield
  {journal} {\bibinfo  {journal} {Physical Review B}\ }\textbf {\bibinfo
  {volume} {81}},\ \bibinfo {pages} {235421} (\bibinfo {year}
  {2010})}\BibitemShut {NoStop}%
\bibitem [{\citenamefont {Zhang}\ \emph {et~al.}(2011)\citenamefont {Zhang},
  \citenamefont {Wang}, \citenamefont {Cheng},\ and\ \citenamefont
  {Xiang}}]{8}%
  \BibitemOpen
  \bibfield  {author} {\bibinfo {author} {\bibfnamefont {Y.}~\bibnamefont
  {Zhang}}, \bibinfo {author} {\bibfnamefont {C.}~\bibnamefont {Wang}},
  \bibinfo {author} {\bibfnamefont {Y.}~\bibnamefont {Cheng}},\ and\ \bibinfo
  {author} {\bibfnamefont {Y.}~\bibnamefont {Xiang}},\ }\bibfield  {title}
  {\bibinfo {title} {Mechanical properties of bilayer graphene sheets coupled
  by sp3 bonding},\ }\href
  {https://doi.org/https://doi.org/10.1016/j.carbon.2011.06.058} {\bibfield
  {journal} {\bibinfo  {journal} {Carbon}\ }\textbf {\bibinfo {volume} {49}},\
  \bibinfo {pages} {4511} (\bibinfo {year} {2011})}\BibitemShut {NoStop}%
\bibitem [{\citenamefont {Nair}\ \emph {et~al.}(2008)\citenamefont {Nair},
  \citenamefont {Blake}, \citenamefont {Grigorenko}, \citenamefont {Novoselov},
  \citenamefont {Booth}, \citenamefont {Stauber}, \citenamefont {Peres},\ and\
  \citenamefont {Geim}}]{9}%
  \BibitemOpen
  \bibfield  {author} {\bibinfo {author} {\bibfnamefont {R.~R.}\ \bibnamefont
  {Nair}}, \bibinfo {author} {\bibfnamefont {P.}~\bibnamefont {Blake}},
  \bibinfo {author} {\bibfnamefont {A.~N.}\ \bibnamefont {Grigorenko}},
  \bibinfo {author} {\bibfnamefont {K.~S.}\ \bibnamefont {Novoselov}}, \bibinfo
  {author} {\bibfnamefont {T.~J.}\ \bibnamefont {Booth}}, \bibinfo {author}
  {\bibfnamefont {T.}~\bibnamefont {Stauber}}, \bibinfo {author} {\bibfnamefont
  {N.~M.}\ \bibnamefont {Peres}},\ and\ \bibinfo {author} {\bibfnamefont
  {A.~K.}\ \bibnamefont {Geim}},\ }\bibfield  {title} {\bibinfo {title} {Fine
  structure constant defines visual transparency of graphene},\ }\href
  {https://doi.org/10.1126/science.11569} {\bibfield  {journal} {\bibinfo
  {journal} {Science}\ }\textbf {\bibinfo {volume} {320}},\ \bibinfo {pages}
  {1308} (\bibinfo {year} {2008})}\BibitemShut {NoStop}%
\bibitem [{\citenamefont {Ohta}\ \emph {et~al.}(2006)\citenamefont {Ohta},
  \citenamefont {Bostwick}, \citenamefont {Seyller}, \citenamefont {Horn},\
  and\ \citenamefont {Rotenberg}}]{10}%
  \BibitemOpen
  \bibfield  {author} {\bibinfo {author} {\bibfnamefont {T.}~\bibnamefont
  {Ohta}}, \bibinfo {author} {\bibfnamefont {A.}~\bibnamefont {Bostwick}},
  \bibinfo {author} {\bibfnamefont {T.}~\bibnamefont {Seyller}}, \bibinfo
  {author} {\bibfnamefont {K.}~\bibnamefont {Horn}},\ and\ \bibinfo {author}
  {\bibfnamefont {E.}~\bibnamefont {Rotenberg}},\ }\bibfield  {title} {\bibinfo
  {title} {Controlling the electronic structure of bilayer graphene},\ }\href
  {https://doi.org/10.1126/science.1130681} {\bibfield  {journal} {\bibinfo
  {journal} {Science}\ }\textbf {\bibinfo {volume} {313}},\ \bibinfo {pages}
  {951} (\bibinfo {year} {2006})}\BibitemShut {NoStop}%
\bibitem [{\citenamefont {McCann}(2006)}]{11}%
  \BibitemOpen
  \bibfield  {author} {\bibinfo {author} {\bibfnamefont {E.}~\bibnamefont
  {McCann}},\ }\bibfield  {title} {\bibinfo {title} {Asymmetry gap in the
  electronic band structure of bilayer graphene},\ }\href
  {https://doi.org/https://doi.org/10.1103/PhysRevB.74.161403} {\bibfield
  {journal} {\bibinfo  {journal} {Physical Review B}\ }\textbf {\bibinfo
  {volume} {74}},\ \bibinfo {pages} {161403} (\bibinfo {year}
  {2006})}\BibitemShut {NoStop}%
\bibitem [{\citenamefont {Castro}\ \emph {et~al.}(2007)\citenamefont {Castro},
  \citenamefont {Novoselov}, \citenamefont {Morozov}, \citenamefont {Peres},
  \citenamefont {Dos~Santos}, \citenamefont {Nilsson}, \citenamefont {Guinea},
  \citenamefont {Geim},\ and\ \citenamefont {Neto}}]{12}%
  \BibitemOpen
  \bibfield  {author} {\bibinfo {author} {\bibfnamefont {E.~V.}\ \bibnamefont
  {Castro}}, \bibinfo {author} {\bibfnamefont {K.}~\bibnamefont {Novoselov}},
  \bibinfo {author} {\bibfnamefont {S.}~\bibnamefont {Morozov}}, \bibinfo
  {author} {\bibfnamefont {N.}~\bibnamefont {Peres}}, \bibinfo {author}
  {\bibfnamefont {J.~L.}\ \bibnamefont {Dos~Santos}}, \bibinfo {author}
  {\bibfnamefont {J.}~\bibnamefont {Nilsson}}, \bibinfo {author} {\bibfnamefont
  {F.}~\bibnamefont {Guinea}}, \bibinfo {author} {\bibfnamefont
  {A.}~\bibnamefont {Geim}},\ and\ \bibinfo {author} {\bibfnamefont {A.~C.}\
  \bibnamefont {Neto}},\ }\bibfield  {title} {\bibinfo {title} {Biased bilayer
  graphene: semiconductor with a gap tunable by the electric field effect},\
  }\href {https://doi.org/https://doi.org/10.1103/PhysRevLett.99.216802}
  {\bibfield  {journal} {\bibinfo  {journal} {Physical review letters}\
  }\textbf {\bibinfo {volume} {99}},\ \bibinfo {pages} {216802} (\bibinfo
  {year} {2007})}\BibitemShut {NoStop}%
\bibitem [{\citenamefont {Oostinga}\ \emph {et~al.}(2008)\citenamefont
  {Oostinga}, \citenamefont {Heersche}, \citenamefont {Liu}, \citenamefont
  {Morpurgo},\ and\ \citenamefont {Vandersypen}}]{13}%
  \BibitemOpen
  \bibfield  {author} {\bibinfo {author} {\bibfnamefont {J.~B.}\ \bibnamefont
  {Oostinga}}, \bibinfo {author} {\bibfnamefont {H.~B.}\ \bibnamefont
  {Heersche}}, \bibinfo {author} {\bibfnamefont {X.}~\bibnamefont {Liu}},
  \bibinfo {author} {\bibfnamefont {A.~F.}\ \bibnamefont {Morpurgo}},\ and\
  \bibinfo {author} {\bibfnamefont {L.~M.}\ \bibnamefont {Vandersypen}},\
  }\bibfield  {title} {\bibinfo {title} {Gate-induced insulating state in
  bilayer graphene devices},\ }\href
  {https://doi.org/https://doi.org/10.1038/nmat2082} {\bibfield  {journal}
  {\bibinfo  {journal} {Nature materials}\ }\textbf {\bibinfo {volume} {7}},\
  \bibinfo {pages} {151} (\bibinfo {year} {2008})}\BibitemShut {NoStop}%
\bibitem [{\citenamefont {Naumis}\ \emph {et~al.}()\citenamefont {Naumis},
  \citenamefont {Herrera}, \citenamefont {Poudel}, \citenamefont {Nakamura},\
  and\ \citenamefont {Barraza-Lopez}}]{naumisstrained}%
  \BibitemOpen
  \bibfield  {author} {\bibinfo {author} {\bibfnamefont {G.~G.}\ \bibnamefont
  {Naumis}}, \bibinfo {author} {\bibfnamefont {S.~A.}\ \bibnamefont {Herrera}},
  \bibinfo {author} {\bibfnamefont {S.~P.}\ \bibnamefont {Poudel}}, \bibinfo
  {author} {\bibfnamefont {H.}~\bibnamefont {Nakamura}},\ and\ \bibinfo
  {author} {\bibfnamefont {S.}~\bibnamefont {Barraza-Lopez}},\ }\bibfield
  {title} {\bibinfo {title} {Mechanical, electronic, optical, piezoelectric and
  ferroic properties of strained graphene and other strained monolayers and
  multilayers: An update},\ }\href {https://doi.org/10.1088/1361-6633/ad06db}
  {\bibfield  {journal} {\bibinfo  {journal} {Rep. Prog. Phys}\ }\textbf
  {\bibinfo {volume} {87}},\ \bibinfo {pages} {016502}}\BibitemShut {NoStop}%
\bibitem [{\citenamefont {Andrei}\ and\ \citenamefont {MacDonald}(2020)}]{14}%
  \BibitemOpen
  \bibfield  {author} {\bibinfo {author} {\bibfnamefont {E.~Y.}\ \bibnamefont
  {Andrei}}\ and\ \bibinfo {author} {\bibfnamefont {A.~H.}\ \bibnamefont
  {MacDonald}},\ }\bibfield  {title} {\bibinfo {title} {Graphene bilayers with
  a twist},\ }\href
  {https://doi.org/https://doi.org/10.1038/s41563-020-00840-0} {\bibfield
  {journal} {\bibinfo  {journal} {Nature materials}\ }\textbf {\bibinfo
  {volume} {19}},\ \bibinfo {pages} {1265} (\bibinfo {year}
  {2020})}\BibitemShut {NoStop}%
\bibitem [{\citenamefont {Brihuega}\ \emph {et~al.}(2012)\citenamefont
  {Brihuega}, \citenamefont {Mallet}, \citenamefont {Gonz{\'a}lez-Herrero},
  \citenamefont {De~Laissardi{\`e}re}, \citenamefont {Ugeda}, \citenamefont
  {Magaud}, \citenamefont {G{\'o}mez-Rodr{\'\i}guez}, \citenamefont
  {Yndur{\'a}in},\ and\ \citenamefont {Veuillen}}]{15}%
  \BibitemOpen
  \bibfield  {author} {\bibinfo {author} {\bibfnamefont {I.}~\bibnamefont
  {Brihuega}}, \bibinfo {author} {\bibfnamefont {P.}~\bibnamefont {Mallet}},
  \bibinfo {author} {\bibfnamefont {H.}~\bibnamefont {Gonz{\'a}lez-Herrero}},
  \bibinfo {author} {\bibfnamefont {G.~T.}\ \bibnamefont
  {De~Laissardi{\`e}re}}, \bibinfo {author} {\bibfnamefont {M.}~\bibnamefont
  {Ugeda}}, \bibinfo {author} {\bibfnamefont {L.}~\bibnamefont {Magaud}},
  \bibinfo {author} {\bibfnamefont {J.}~\bibnamefont
  {G{\'o}mez-Rodr{\'\i}guez}}, \bibinfo {author} {\bibfnamefont
  {F.}~\bibnamefont {Yndur{\'a}in}},\ and\ \bibinfo {author} {\bibfnamefont
  {J.-Y.}\ \bibnamefont {Veuillen}},\ }\bibfield  {title} {\bibinfo {title}
  {Unraveling the intrinsic and robust nature of van hove singularities in
  twisted bilayer graphene by scanning tunneling microscopy and theoretical
  analysis},\ }\href
  {https://doi.org/https://doi.org/10.1103/PhysRevLett.109.196802} {\bibfield
  {journal} {\bibinfo  {journal} {Physical review letters}\ }\textbf {\bibinfo
  {volume} {109}},\ \bibinfo {pages} {196802} (\bibinfo {year}
  {2012})}\BibitemShut {NoStop}%
\bibitem [{\citenamefont {Cisternas}\ and\ \citenamefont {Correa}(2012)}]{16}%
  \BibitemOpen
  \bibfield  {author} {\bibinfo {author} {\bibfnamefont {E.}~\bibnamefont
  {Cisternas}}\ and\ \bibinfo {author} {\bibfnamefont {J.}~\bibnamefont
  {Correa}},\ }\bibfield  {title} {\bibinfo {title} {Theoretical reproduction
  of superstructures revealed by stm on bilayer graphene},\ }\href
  {https://doi.org/https://doi.org/10.1016/j.chemphys.2012.09.021} {\bibfield
  {journal} {\bibinfo  {journal} {Chemical Physics}\ }\textbf {\bibinfo
  {volume} {409}},\ \bibinfo {pages} {74} (\bibinfo {year} {2012})}\BibitemShut
  {NoStop}%
\bibitem [{\citenamefont {Varchon}\ \emph {et~al.}(2008)\citenamefont
  {Varchon}, \citenamefont {Mallet}, \citenamefont {Magaud},\ and\
  \citenamefont {Veuillen}}]{TBLG2}%
  \BibitemOpen
  \bibfield  {author} {\bibinfo {author} {\bibfnamefont {F.}~\bibnamefont
  {Varchon}}, \bibinfo {author} {\bibfnamefont {P.}~\bibnamefont {Mallet}},
  \bibinfo {author} {\bibfnamefont {L.}~\bibnamefont {Magaud}},\ and\ \bibinfo
  {author} {\bibfnamefont {J.-Y.}\ \bibnamefont {Veuillen}},\ }\bibfield
  {title} {\bibinfo {title} {Rotational disorder in few-layer graphene films on
  6 h- si c (000- 1): A scanning tunneling microscopy study},\ }\href
  {https://doi.org/https://doi.org/10.1103/PhysRevB.77.165415} {\bibfield
  {journal} {\bibinfo  {journal} {Physical Review B}\ }\textbf {\bibinfo
  {volume} {77}},\ \bibinfo {pages} {165415} (\bibinfo {year}
  {2008})}\BibitemShut {NoStop}%
\bibitem [{\citenamefont {Cao}\ \emph {et~al.}(2018{\natexlab{a}})\citenamefont
  {Cao}, \citenamefont {Fatemi}, \citenamefont {Demir}, \citenamefont {Fang},
  \citenamefont {Tomarken}, \citenamefont {Luo}, \citenamefont
  {Sanchez-Yamagishi}, \citenamefont {Watanabe}, \citenamefont {Taniguchi},
  \citenamefont {Kaxiras} \emph {et~al.}}]{17}%
  \BibitemOpen
  \bibfield  {author} {\bibinfo {author} {\bibfnamefont {Y.}~\bibnamefont
  {Cao}}, \bibinfo {author} {\bibfnamefont {V.}~\bibnamefont {Fatemi}},
  \bibinfo {author} {\bibfnamefont {A.}~\bibnamefont {Demir}}, \bibinfo
  {author} {\bibfnamefont {S.}~\bibnamefont {Fang}}, \bibinfo {author}
  {\bibfnamefont {S.~L.}\ \bibnamefont {Tomarken}}, \bibinfo {author}
  {\bibfnamefont {J.~Y.}\ \bibnamefont {Luo}}, \bibinfo {author} {\bibfnamefont
  {J.~D.}\ \bibnamefont {Sanchez-Yamagishi}}, \bibinfo {author} {\bibfnamefont
  {K.}~\bibnamefont {Watanabe}}, \bibinfo {author} {\bibfnamefont
  {T.}~\bibnamefont {Taniguchi}}, \bibinfo {author} {\bibfnamefont
  {E.}~\bibnamefont {Kaxiras}}, \emph {et~al.},\ }\bibfield  {title} {\bibinfo
  {title} {Correlated insulator behaviour at half-filling in magic-angle
  graphene superlattices},\ }\href
  {https://doi.org/https://doi.org/10.1038/nature26154} {\bibfield  {journal}
  {\bibinfo  {journal} {Nature}\ }\textbf {\bibinfo {volume} {556}},\ \bibinfo
  {pages} {80} (\bibinfo {year} {2018}{\natexlab{a}})}\BibitemShut {NoStop}%
\bibitem [{\citenamefont {Cao}\ \emph {et~al.}(2018{\natexlab{b}})\citenamefont
  {Cao}, \citenamefont {Fatemi}, \citenamefont {Fang}, \citenamefont
  {Watanabe}, \citenamefont {Taniguchi}, \citenamefont {Kaxiras},\ and\
  \citenamefont {Jarillo-Herrero}}]{18}%
  \BibitemOpen
  \bibfield  {author} {\bibinfo {author} {\bibfnamefont {Y.}~\bibnamefont
  {Cao}}, \bibinfo {author} {\bibfnamefont {V.}~\bibnamefont {Fatemi}},
  \bibinfo {author} {\bibfnamefont {S.}~\bibnamefont {Fang}}, \bibinfo {author}
  {\bibfnamefont {K.}~\bibnamefont {Watanabe}}, \bibinfo {author}
  {\bibfnamefont {T.}~\bibnamefont {Taniguchi}}, \bibinfo {author}
  {\bibfnamefont {E.}~\bibnamefont {Kaxiras}},\ and\ \bibinfo {author}
  {\bibfnamefont {P.}~\bibnamefont {Jarillo-Herrero}},\ }\bibfield  {title}
  {\bibinfo {title} {Unconventional superconductivity in magic-angle graphene
  superlattices},\ }\href {https://doi.org/10.1038/nature26160} {\bibfield
  {journal} {\bibinfo  {journal} {Nature}\ }\textbf {\bibinfo {volume} {556}},\
  \bibinfo {pages} {43} (\bibinfo {year} {2018}{\natexlab{b}})}\BibitemShut
  {NoStop}%
\bibitem [{\citenamefont {Lau}\ \emph {et~al.}(2022)\citenamefont {Lau},
  \citenamefont {Bockrath}, \citenamefont {Mak},\ and\ \citenamefont
  {Zhang}}]{TBLG3}%
  \BibitemOpen
  \bibfield  {author} {\bibinfo {author} {\bibfnamefont {C.~N.}\ \bibnamefont
  {Lau}}, \bibinfo {author} {\bibfnamefont {M.~W.}\ \bibnamefont {Bockrath}},
  \bibinfo {author} {\bibfnamefont {K.~F.}\ \bibnamefont {Mak}},\ and\ \bibinfo
  {author} {\bibfnamefont {F.}~\bibnamefont {Zhang}},\ }\bibfield  {title}
  {\bibinfo {title} {Reproducibility in the fabrication and physics of
  moir{\'e} materials},\ }\href
  {https://doi.org/https://doi.org/10.1038/s41586-021-04173-z} {\bibfield
  {journal} {\bibinfo  {journal} {Nature}\ }\textbf {\bibinfo {volume} {602}},\
  \bibinfo {pages} {41} (\bibinfo {year} {2022})}\BibitemShut {NoStop}%
\bibitem [{\citenamefont {Gargiulo}\ and\ \citenamefont {Yazyev}(2017)}]{25}%
  \BibitemOpen
  \bibfield  {author} {\bibinfo {author} {\bibfnamefont {F.}~\bibnamefont
  {Gargiulo}}\ and\ \bibinfo {author} {\bibfnamefont {O.~V.}\ \bibnamefont
  {Yazyev}},\ }\bibfield  {title} {\bibinfo {title} {Structural and electronic
  transformation in low-angle twisted bilayer graphene},\ }\href
  {https://doi.org/10.1088/2053-1583/aa9640} {\bibfield  {journal} {\bibinfo
  {journal} {2D Materials}\ }\textbf {\bibinfo {volume} {5}},\ \bibinfo {pages}
  {015019} (\bibinfo {year} {2017})}\BibitemShut {NoStop}%
\bibitem [{\citenamefont {Uchida}\ \emph {et~al.}(2014)\citenamefont {Uchida},
  \citenamefont {Furuya}, \citenamefont {Iwata},\ and\ \citenamefont
  {Oshiyama}}]{26}%
  \BibitemOpen
  \bibfield  {author} {\bibinfo {author} {\bibfnamefont {K.}~\bibnamefont
  {Uchida}}, \bibinfo {author} {\bibfnamefont {S.}~\bibnamefont {Furuya}},
  \bibinfo {author} {\bibfnamefont {J.-I.}\ \bibnamefont {Iwata}},\ and\
  \bibinfo {author} {\bibfnamefont {A.}~\bibnamefont {Oshiyama}},\ }\bibfield
  {title} {\bibinfo {title} {Atomic corrugation and electron localization due
  to moir{\'e} patterns in twisted bilayer graphenes},\ }\href
  {https://doi.org/https://doi.org/10.1103/PhysRevB.90.155451} {\bibfield
  {journal} {\bibinfo  {journal} {Physical Review B}\ }\textbf {\bibinfo
  {volume} {90}},\ \bibinfo {pages} {155451} (\bibinfo {year}
  {2014})}\BibitemShut {NoStop}%
\bibitem [{\citenamefont {Jain}\ \emph {et~al.}(2016)\citenamefont {Jain},
  \citenamefont {Juričić},\ and\ \citenamefont {Barkema}}]{buckle1}%
  \BibitemOpen
  \bibfield  {author} {\bibinfo {author} {\bibfnamefont {S.~K.}\ \bibnamefont
  {Jain}}, \bibinfo {author} {\bibfnamefont {V.}~\bibnamefont {Juričić}},\
  and\ \bibinfo {author} {\bibfnamefont {G.~T.}\ \bibnamefont {Barkema}},\
  }\bibfield  {title} {\bibinfo {title} {Structure of twisted and buckled
  bilayer graphene},\ }\href {https://doi.org/10.1088/2053-1583/4/1/015018}
  {\bibfield  {journal} {\bibinfo  {journal} {2D Materials}\ }\textbf {\bibinfo
  {volume} {4}},\ \bibinfo {pages} {015018} (\bibinfo {year}
  {2016})}\BibitemShut {NoStop}%
\bibitem [{\citenamefont {Liu}\ \emph {et~al.}(2021)\citenamefont {Liu},
  \citenamefont {Xian}, \citenamefont {Mu}, \citenamefont {Zhao}, \citenamefont
  {Liu}, \citenamefont {Rubio},\ and\ \citenamefont {Wang}}]{gaps1}%
  \BibitemOpen
  \bibfield  {author} {\bibinfo {author} {\bibfnamefont {B.}~\bibnamefont
  {Liu}}, \bibinfo {author} {\bibfnamefont {L.}~\bibnamefont {Xian}}, \bibinfo
  {author} {\bibfnamefont {H.}~\bibnamefont {Mu}}, \bibinfo {author}
  {\bibfnamefont {G.}~\bibnamefont {Zhao}}, \bibinfo {author} {\bibfnamefont
  {Z.}~\bibnamefont {Liu}}, \bibinfo {author} {\bibfnamefont {A.}~\bibnamefont
  {Rubio}},\ and\ \bibinfo {author} {\bibfnamefont {Z.}~\bibnamefont {Wang}},\
  }\bibfield  {title} {\bibinfo {title} {Higher-order band topology in twisted
  moir{\'e} superlattice},\ }\href
  {https://doi.org/https://doi.org/10.1103/PhysRevLett.126.066401} {\bibfield
  {journal} {\bibinfo  {journal} {Physical Review Letters}\ }\textbf {\bibinfo
  {volume} {126}},\ \bibinfo {pages} {066401} (\bibinfo {year}
  {2021})}\BibitemShut {NoStop}%
\bibitem [{\citenamefont {Fortin-Desch{\^e}nes}\ \emph
  {et~al.}(2022)\citenamefont {Fortin-Desch{\^e}nes}, \citenamefont {Pu},
  \citenamefont {Zhou}, \citenamefont {Ma}, \citenamefont {Cheung},
  \citenamefont {Watanabe}, \citenamefont {Taniguchi}, \citenamefont {Zhang},
  \citenamefont {Du},\ and\ \citenamefont {Xia}}]{gaps2}%
  \BibitemOpen
  \bibfield  {author} {\bibinfo {author} {\bibfnamefont {M.}~\bibnamefont
  {Fortin-Desch{\^e}nes}}, \bibinfo {author} {\bibfnamefont {R.}~\bibnamefont
  {Pu}}, \bibinfo {author} {\bibfnamefont {Y.-F.}\ \bibnamefont {Zhou}},
  \bibinfo {author} {\bibfnamefont {C.}~\bibnamefont {Ma}}, \bibinfo {author}
  {\bibfnamefont {P.}~\bibnamefont {Cheung}}, \bibinfo {author} {\bibfnamefont
  {K.}~\bibnamefont {Watanabe}}, \bibinfo {author} {\bibfnamefont
  {T.}~\bibnamefont {Taniguchi}}, \bibinfo {author} {\bibfnamefont
  {F.}~\bibnamefont {Zhang}}, \bibinfo {author} {\bibfnamefont
  {X.}~\bibnamefont {Du}},\ and\ \bibinfo {author} {\bibfnamefont
  {F.}~\bibnamefont {Xia}},\ }\bibfield  {title} {\bibinfo {title} {Uncovering
  topological edge states in twisted bilayer graphene},\ }\href
  {https://doi.org/https://doi.org/10.1021/acs.nanolett.2c01481} {\bibfield
  {journal} {\bibinfo  {journal} {Nano Letters}\ }\textbf {\bibinfo {volume}
  {22}},\ \bibinfo {pages} {6186} (\bibinfo {year} {2022})}\BibitemShut
  {NoStop}%
\bibitem [{\citenamefont {Dos~Santos}\ \emph {et~al.}(2007)\citenamefont
  {Dos~Santos}, \citenamefont {Peres},\ and\ \citenamefont {Neto}}]{19}%
  \BibitemOpen
  \bibfield  {author} {\bibinfo {author} {\bibfnamefont {J.~L.}\ \bibnamefont
  {Dos~Santos}}, \bibinfo {author} {\bibfnamefont {N.}~\bibnamefont {Peres}},\
  and\ \bibinfo {author} {\bibfnamefont {A.~C.}\ \bibnamefont {Neto}},\
  }\bibfield  {title} {\bibinfo {title} {Graphene bilayer with a twist:
  electronic structure},\ }\href
  {https://doi.org/https://doi.org/10.1103/PhysRevLett.99.256802} {\bibfield
  {journal} {\bibinfo  {journal} {Physical review letters}\ }\textbf {\bibinfo
  {volume} {99}},\ \bibinfo {pages} {256802} (\bibinfo {year}
  {2007})}\BibitemShut {NoStop}%
\bibitem [{\citenamefont {Morell}\ \emph {et~al.}(2010)\citenamefont {Morell},
  \citenamefont {Correa}, \citenamefont {Vargas}, \citenamefont {Pacheco},\
  and\ \citenamefont {Barticevic}}]{20}%
  \BibitemOpen
  \bibfield  {author} {\bibinfo {author} {\bibfnamefont {E.~S.}\ \bibnamefont
  {Morell}}, \bibinfo {author} {\bibfnamefont {J.}~\bibnamefont {Correa}},
  \bibinfo {author} {\bibfnamefont {P.}~\bibnamefont {Vargas}}, \bibinfo
  {author} {\bibfnamefont {M.}~\bibnamefont {Pacheco}},\ and\ \bibinfo {author}
  {\bibfnamefont {Z.}~\bibnamefont {Barticevic}},\ }\bibfield  {title}
  {\bibinfo {title} {Flat bands in slightly twisted bilayer graphene:
  Tight-binding calculations},\ }\href
  {https://doi.org/https://doi.org/10.1103/PhysRevB.82.121407} {\bibfield
  {journal} {\bibinfo  {journal} {Physical Review B}\ }\textbf {\bibinfo
  {volume} {82}},\ \bibinfo {pages} {121407} (\bibinfo {year}
  {2010})}\BibitemShut {NoStop}%
\bibitem [{\citenamefont {Hass}\ \emph {et~al.}(2008)\citenamefont {Hass},
  \citenamefont {Varchon}, \citenamefont {Millan-Otoya}, \citenamefont
  {Sprinkle}, \citenamefont {Sharma}, \citenamefont {de~Heer}, \citenamefont
  {Berger}, \citenamefont {First}, \citenamefont {Magaud},\ and\ \citenamefont
  {Conrad}}]{TBLG1}%
  \BibitemOpen
  \bibfield  {author} {\bibinfo {author} {\bibfnamefont {J.}~\bibnamefont
  {Hass}}, \bibinfo {author} {\bibfnamefont {F.}~\bibnamefont {Varchon}},
  \bibinfo {author} {\bibfnamefont {J.-E.}\ \bibnamefont {Millan-Otoya}},
  \bibinfo {author} {\bibfnamefont {M.}~\bibnamefont {Sprinkle}}, \bibinfo
  {author} {\bibfnamefont {N.}~\bibnamefont {Sharma}}, \bibinfo {author}
  {\bibfnamefont {W.~A.}\ \bibnamefont {de~Heer}}, \bibinfo {author}
  {\bibfnamefont {C.}~\bibnamefont {Berger}}, \bibinfo {author} {\bibfnamefont
  {P.~N.}\ \bibnamefont {First}}, \bibinfo {author} {\bibfnamefont
  {L.}~\bibnamefont {Magaud}},\ and\ \bibinfo {author} {\bibfnamefont {E.~H.}\
  \bibnamefont {Conrad}},\ }\bibfield  {title} {\bibinfo {title} {Why
  multilayer graphene on 4 h- sic (000 1) behaves like a single sheet of
  graphene},\ }\href
  {https://doi.org/https://doi.org/10.1103/PhysRevLett.100.125504} {\bibfield
  {journal} {\bibinfo  {journal} {Physical review letters}\ }\textbf {\bibinfo
  {volume} {100}},\ \bibinfo {pages} {125504} (\bibinfo {year}
  {2008})}\BibitemShut {NoStop}%
\bibitem [{\citenamefont {Moon}\ and\ \citenamefont {Koshino}(2013)}]{21}%
  \BibitemOpen
  \bibfield  {author} {\bibinfo {author} {\bibfnamefont {P.}~\bibnamefont
  {Moon}}\ and\ \bibinfo {author} {\bibfnamefont {M.}~\bibnamefont {Koshino}},\
  }\bibfield  {title} {\bibinfo {title} {Optical absorption in twisted bilayer
  graphene},\ }\href
  {https://doi.org/https://doi.org/10.1103/PhysRevB.87.205404} {\bibfield
  {journal} {\bibinfo  {journal} {Physical Review B}\ }\textbf {\bibinfo
  {volume} {87}},\ \bibinfo {pages} {205404} (\bibinfo {year}
  {2013})}\BibitemShut {NoStop}%
\bibitem [{\citenamefont {Li}\ \emph {et~al.}(2010)\citenamefont {Li},
  \citenamefont {Luican}, \citenamefont {Lopes~dos Santos}, \citenamefont
  {Castro~Neto}, \citenamefont {Reina}, \citenamefont {Kong},\ and\
  \citenamefont {Andrei}}]{22}%
  \BibitemOpen
  \bibfield  {author} {\bibinfo {author} {\bibfnamefont {G.}~\bibnamefont
  {Li}}, \bibinfo {author} {\bibfnamefont {A.}~\bibnamefont {Luican}}, \bibinfo
  {author} {\bibfnamefont {J.}~\bibnamefont {Lopes~dos Santos}}, \bibinfo
  {author} {\bibfnamefont {A.}~\bibnamefont {Castro~Neto}}, \bibinfo {author}
  {\bibfnamefont {A.}~\bibnamefont {Reina}}, \bibinfo {author} {\bibfnamefont
  {J.}~\bibnamefont {Kong}},\ and\ \bibinfo {author} {\bibfnamefont
  {E.}~\bibnamefont {Andrei}},\ }\bibfield  {title} {\bibinfo {title}
  {Observation of van hove singularities in twisted graphene layers},\ }\href
  {https://doi.org/https://doi.org/10.1038/nphys1463} {\bibfield  {journal}
  {\bibinfo  {journal} {Nature physics}\ }\textbf {\bibinfo {volume} {6}},\
  \bibinfo {pages} {109} (\bibinfo {year} {2010})}\BibitemShut {NoStop}%
\bibitem [{\citenamefont {Bistritzer}\ and\ \citenamefont
  {MacDonald}(2011)}]{macdonald}%
  \BibitemOpen
  \bibfield  {author} {\bibinfo {author} {\bibfnamefont {R.}~\bibnamefont
  {Bistritzer}}\ and\ \bibinfo {author} {\bibfnamefont {A.~H.}\ \bibnamefont
  {MacDonald}},\ }\bibfield  {title} {\bibinfo {title} {Moir{\'e} bands in
  twisted double-layer graphene},\ }\href
  {https://doi.org/https://doi.org/10.1073/pnas.1108174108} {\bibfield
  {journal} {\bibinfo  {journal} {Proceedings of the National Academy of
  Sciences}\ }\textbf {\bibinfo {volume} {108}},\ \bibinfo {pages} {12233}
  (\bibinfo {year} {2011})}\BibitemShut {NoStop}%
\bibitem [{\citenamefont {Naumis}\ \emph {et~al.}(2021)\citenamefont {Naumis},
  \citenamefont {Navarro-Labastida}, \citenamefont {Aguilar-M{\'e}ndez},\ and\
  \citenamefont {Espinosa-Champo}}]{naumis2x2}%
  \BibitemOpen
  \bibfield  {author} {\bibinfo {author} {\bibfnamefont {G.~G.}\ \bibnamefont
  {Naumis}}, \bibinfo {author} {\bibfnamefont {L.~A.}\ \bibnamefont
  {Navarro-Labastida}}, \bibinfo {author} {\bibfnamefont {E.}~\bibnamefont
  {Aguilar-M{\'e}ndez}},\ and\ \bibinfo {author} {\bibfnamefont
  {A.}~\bibnamefont {Espinosa-Champo}},\ }\bibfield  {title} {\bibinfo {title}
  {Reduction of the twisted bilayer graphene chiral hamiltonian into a
  2$\times$ 2 matrix operator and physical origin of flat bands at magic
  angles},\ }\href
  {https://doi.org/https://doi.org/10.1103/PhysRevB.103.245418} {\bibfield
  {journal} {\bibinfo  {journal} {Physical Review B}\ }\textbf {\bibinfo
  {volume} {103}},\ \bibinfo {pages} {245418} (\bibinfo {year}
  {2021})}\BibitemShut {NoStop}%
\bibitem [{\citenamefont {Navarro-Labastida}\ \emph {et~al.}(2022)\citenamefont
  {Navarro-Labastida}, \citenamefont {Espinosa-Champo}, \citenamefont
  {Aguilar-Mendez},\ and\ \citenamefont {Naumis}}]{MATBLG}%
  \BibitemOpen
  \bibfield  {author} {\bibinfo {author} {\bibfnamefont {L.~A.}\ \bibnamefont
  {Navarro-Labastida}}, \bibinfo {author} {\bibfnamefont {A.}~\bibnamefont
  {Espinosa-Champo}}, \bibinfo {author} {\bibfnamefont {E.}~\bibnamefont
  {Aguilar-Mendez}},\ and\ \bibinfo {author} {\bibfnamefont {G.~G.}\
  \bibnamefont {Naumis}},\ }\bibfield  {title} {\bibinfo {title} {Why the first
  magic-angle is different from others in twisted graphene bilayers: Interlayer
  currents, kinetic and confinement energy, and wave-function localization},\
  }\href {https://doi.org/https://doi.org/10.1103/PhysRevB.105.115434}
  {\bibfield  {journal} {\bibinfo  {journal} {Physical Review B}\ }\textbf
  {\bibinfo {volume} {105}},\ \bibinfo {pages} {115434} (\bibinfo {year}
  {2022})}\BibitemShut {NoStop}%
\bibitem [{\citenamefont {Lisi}\ \emph {et~al.}(2021)\citenamefont {Lisi},
  \citenamefont {Lu}, \citenamefont {Benschop}, \citenamefont {de~Jong},
  \citenamefont {Stepanov}, \citenamefont {Duran}, \citenamefont {Margot},
  \citenamefont {Cucchi}, \citenamefont {Cappelli}, \citenamefont {Hunter}
  \emph {et~al.}}]{23}%
  \BibitemOpen
  \bibfield  {author} {\bibinfo {author} {\bibfnamefont {S.}~\bibnamefont
  {Lisi}}, \bibinfo {author} {\bibfnamefont {X.}~\bibnamefont {Lu}}, \bibinfo
  {author} {\bibfnamefont {T.}~\bibnamefont {Benschop}}, \bibinfo {author}
  {\bibfnamefont {T.~A.}\ \bibnamefont {de~Jong}}, \bibinfo {author}
  {\bibfnamefont {P.}~\bibnamefont {Stepanov}}, \bibinfo {author}
  {\bibfnamefont {J.~R.}\ \bibnamefont {Duran}}, \bibinfo {author}
  {\bibfnamefont {F.}~\bibnamefont {Margot}}, \bibinfo {author} {\bibfnamefont
  {I.}~\bibnamefont {Cucchi}}, \bibinfo {author} {\bibfnamefont
  {E.}~\bibnamefont {Cappelli}}, \bibinfo {author} {\bibfnamefont
  {A.}~\bibnamefont {Hunter}}, \emph {et~al.},\ }\bibfield  {title} {\bibinfo
  {title} {Observation of flat bands in twisted bilayer graphene},\ }\href
  {https://doi.org/https://doi.org/10.1038/s41567-020-01041-x} {\bibfield
  {journal} {\bibinfo  {journal} {Nature Physics}\ }\textbf {\bibinfo {volume}
  {17}},\ \bibinfo {pages} {189} (\bibinfo {year} {2021})}\BibitemShut
  {NoStop}%
\bibitem [{\citenamefont {S{\'a}nchez-S{\'a}nchez}\ \emph
  {et~al.}(2022)\citenamefont {S{\'a}nchez-S{\'a}nchez}, \citenamefont
  {Navarro-Espino}, \citenamefont {Betancur-Ocampo}, \citenamefont
  {Barrios-Vargas},\ and\ \citenamefont {Stegmann}}]{nosotros}%
  \BibitemOpen
  \bibfield  {author} {\bibinfo {author} {\bibfnamefont {J.~A.}\ \bibnamefont
  {S{\'a}nchez-S{\'a}nchez}}, \bibinfo {author} {\bibfnamefont
  {M.}~\bibnamefont {Navarro-Espino}}, \bibinfo {author} {\bibfnamefont
  {Y.}~\bibnamefont {Betancur-Ocampo}}, \bibinfo {author} {\bibfnamefont
  {J.~E.}\ \bibnamefont {Barrios-Vargas}},\ and\ \bibinfo {author}
  {\bibfnamefont {T.}~\bibnamefont {Stegmann}},\ }\bibfield  {title} {\bibinfo
  {title} {Steering the current flow in twisted bilayer graphene},\ }\href
  {https://doi.org/10.1088/2515-7639/ac4ae0} {\bibfield  {journal} {\bibinfo
  {journal} {Journal of Physics: Materials}\ }\textbf {\bibinfo {volume} {5}},\
  \bibinfo {pages} {024003} (\bibinfo {year} {2022})}\BibitemShut {NoStop}%
\bibitem [{\citenamefont {Sainz-Cruz}\ \emph {et~al.}(2021)\citenamefont
  {Sainz-Cruz}, \citenamefont {Cea}, \citenamefont {Pantale{\'o}n},\ and\
  \citenamefont {Guinea}}]{stuff1}%
  \BibitemOpen
  \bibfield  {author} {\bibinfo {author} {\bibfnamefont {H.}~\bibnamefont
  {Sainz-Cruz}}, \bibinfo {author} {\bibfnamefont {T.}~\bibnamefont {Cea}},
  \bibinfo {author} {\bibfnamefont {P.~A.}\ \bibnamefont {Pantale{\'o}n}},\
  and\ \bibinfo {author} {\bibfnamefont {F.}~\bibnamefont {Guinea}},\
  }\bibfield  {title} {\bibinfo {title} {High transmission in twisted bilayer
  graphene with angle disorder},\ }\href
  {https://doi.org/https://doi.org/10.1103/PhysRevB.104.075144} {\bibfield
  {journal} {\bibinfo  {journal} {Physical Review B}\ }\textbf {\bibinfo
  {volume} {104}},\ \bibinfo {pages} {075144} (\bibinfo {year}
  {2021})}\BibitemShut {NoStop}%
\bibitem [{\citenamefont {Wang}\ \emph {et~al.}(2023)\citenamefont {Wang},
  \citenamefont {Bao}, \citenamefont {Zheng}, \citenamefont {Wang},
  \citenamefont {Wang}, \citenamefont {Fan}, \citenamefont {Mishra},
  \citenamefont {Tao}, \citenamefont {Xiao}, \citenamefont {Huang} \emph
  {et~al.}}]{edge2}%
  \BibitemOpen
  \bibfield  {author} {\bibinfo {author} {\bibfnamefont {D.}~\bibnamefont
  {Wang}}, \bibinfo {author} {\bibfnamefont {D.-L.}\ \bibnamefont {Bao}},
  \bibinfo {author} {\bibfnamefont {Q.}~\bibnamefont {Zheng}}, \bibinfo
  {author} {\bibfnamefont {C.-T.}\ \bibnamefont {Wang}}, \bibinfo {author}
  {\bibfnamefont {S.}~\bibnamefont {Wang}}, \bibinfo {author} {\bibfnamefont
  {P.}~\bibnamefont {Fan}}, \bibinfo {author} {\bibfnamefont {S.}~\bibnamefont
  {Mishra}}, \bibinfo {author} {\bibfnamefont {L.}~\bibnamefont {Tao}},
  \bibinfo {author} {\bibfnamefont {Y.}~\bibnamefont {Xiao}}, \bibinfo {author}
  {\bibfnamefont {L.}~\bibnamefont {Huang}}, \emph {et~al.},\ }\bibfield
  {title} {\bibinfo {title} {Twisted bilayer zigzag-graphene nanoribbon
  junctions with tunable edge states},\ }\href
  {https://doi.org/https://doi.org/10.1038/s41467-023-36613-x} {\bibfield
  {journal} {\bibinfo  {journal} {Nature Communications}\ }\textbf {\bibinfo
  {volume} {14}},\ \bibinfo {pages} {1018} (\bibinfo {year}
  {2023})}\BibitemShut {NoStop}%
\bibitem [{\citenamefont {Sharpe}\ \emph {et~al.}(2019)\citenamefont {Sharpe},
  \citenamefont {Fox}, \citenamefont {Barnard}, \citenamefont {Finney},
  \citenamefont {Watanabe}, \citenamefont {Taniguchi}, \citenamefont
  {Kastner},\ and\ \citenamefont {Goldhaber-Gordon}}]{edge1}%
  \BibitemOpen
  \bibfield  {author} {\bibinfo {author} {\bibfnamefont {A.~L.}\ \bibnamefont
  {Sharpe}}, \bibinfo {author} {\bibfnamefont {E.~J.}\ \bibnamefont {Fox}},
  \bibinfo {author} {\bibfnamefont {A.~W.}\ \bibnamefont {Barnard}}, \bibinfo
  {author} {\bibfnamefont {J.}~\bibnamefont {Finney}}, \bibinfo {author}
  {\bibfnamefont {K.}~\bibnamefont {Watanabe}}, \bibinfo {author}
  {\bibfnamefont {T.}~\bibnamefont {Taniguchi}}, \bibinfo {author}
  {\bibfnamefont {M.}~\bibnamefont {Kastner}},\ and\ \bibinfo {author}
  {\bibfnamefont {D.}~\bibnamefont {Goldhaber-Gordon}},\ }\bibfield  {title}
  {\bibinfo {title} {Emergent ferromagnetism near three-quarters filling in
  twisted bilayer graphene},\ }\href {https://doi.org/10.1126/science.aaw3780}
  {\bibfield  {journal} {\bibinfo  {journal} {Science}\ }\textbf {\bibinfo
  {volume} {365}},\ \bibinfo {pages} {605} (\bibinfo {year}
  {2019})}\BibitemShut {NoStop}%
\bibitem [{\citenamefont {Fleischmann}\ \emph {et~al.}(2018)\citenamefont
  {Fleischmann}, \citenamefont {Gupta}, \citenamefont {Weckbecker},
  \citenamefont {Landgraf}, \citenamefont {Pankratov}, \citenamefont {Meded},\
  and\ \citenamefont {Shallcross}}]{edge3}%
  \BibitemOpen
  \bibfield  {author} {\bibinfo {author} {\bibfnamefont {M.}~\bibnamefont
  {Fleischmann}}, \bibinfo {author} {\bibfnamefont {R.}~\bibnamefont {Gupta}},
  \bibinfo {author} {\bibfnamefont {D.}~\bibnamefont {Weckbecker}}, \bibinfo
  {author} {\bibfnamefont {W.}~\bibnamefont {Landgraf}}, \bibinfo {author}
  {\bibfnamefont {O.}~\bibnamefont {Pankratov}}, \bibinfo {author}
  {\bibfnamefont {V.}~\bibnamefont {Meded}},\ and\ \bibinfo {author}
  {\bibfnamefont {S.}~\bibnamefont {Shallcross}},\ }\bibfield  {title}
  {\bibinfo {title} {Moir{\'e} edge states in twisted graphene nanoribbons},\
  }\href {https://doi.org/https://doi.org/10.1103/PhysRevB.97.205128}
  {\bibfield  {journal} {\bibinfo  {journal} {Physical Review B}\ }\textbf
  {\bibinfo {volume} {97}},\ \bibinfo {pages} {205128} (\bibinfo {year}
  {2018})}\BibitemShut {NoStop}%
\bibitem [{\citenamefont {Fujimoto}\ and\ \citenamefont
  {Koshino}(2021)}]{slide}%
  \BibitemOpen
  \bibfield  {author} {\bibinfo {author} {\bibfnamefont {M.}~\bibnamefont
  {Fujimoto}}\ and\ \bibinfo {author} {\bibfnamefont {M.}~\bibnamefont
  {Koshino}},\ }\bibfield  {title} {\bibinfo {title} {Moir{\'e} edge states in
  twisted bilayer graphene and their topological relation to quantum pumping},\
  }\href {https://doi.org/https://doi.org/10.1103/PhysRevB.103.155410}
  {\bibfield  {journal} {\bibinfo  {journal} {Physical Review B}\ }\textbf
  {\bibinfo {volume} {103}},\ \bibinfo {pages} {155410} (\bibinfo {year}
  {2021})}\BibitemShut {NoStop}%
\bibitem [{\citenamefont {Ma}\ \emph {et~al.}(2020)\citenamefont {Ma},
  \citenamefont {Wang}, \citenamefont {Mills}, \citenamefont {Chen},
  \citenamefont {Deng}, \citenamefont {Yuan}, \citenamefont {Li}, \citenamefont
  {Watanabe}, \citenamefont {Taniguchi}, \citenamefont {Du} \emph
  {et~al.}}]{24}%
  \BibitemOpen
  \bibfield  {author} {\bibinfo {author} {\bibfnamefont {C.}~\bibnamefont
  {Ma}}, \bibinfo {author} {\bibfnamefont {Q.}~\bibnamefont {Wang}}, \bibinfo
  {author} {\bibfnamefont {S.}~\bibnamefont {Mills}}, \bibinfo {author}
  {\bibfnamefont {X.}~\bibnamefont {Chen}}, \bibinfo {author} {\bibfnamefont
  {B.}~\bibnamefont {Deng}}, \bibinfo {author} {\bibfnamefont {S.}~\bibnamefont
  {Yuan}}, \bibinfo {author} {\bibfnamefont {C.}~\bibnamefont {Li}}, \bibinfo
  {author} {\bibfnamefont {K.}~\bibnamefont {Watanabe}}, \bibinfo {author}
  {\bibfnamefont {T.}~\bibnamefont {Taniguchi}}, \bibinfo {author}
  {\bibfnamefont {X.}~\bibnamefont {Du}}, \emph {et~al.},\ }\bibfield  {title}
  {\bibinfo {title} {Moir{\'e} band topology in twisted bilayer graphene},\
  }\href {https://doi.org/https://doi.org/10.1021/acs.nanolett.0c02131}
  {\bibfield  {journal} {\bibinfo  {journal} {Nano letters}\ }\textbf {\bibinfo
  {volume} {20}},\ \bibinfo {pages} {6076} (\bibinfo {year}
  {2020})}\BibitemShut {NoStop}%
\bibitem [{\citenamefont {Zhang}\ \emph {et~al.}(2020)\citenamefont {Zhang},
  \citenamefont {Song}, \citenamefont {Chen}, \citenamefont {Jiang},
  \citenamefont {Chen}, \citenamefont {Gao}, \citenamefont {Hou}, \citenamefont
  {Liu}, \citenamefont {Ma}, \citenamefont {Wang} \emph {et~al.}}]{dft1}%
  \BibitemOpen
  \bibfield  {author} {\bibinfo {author} {\bibfnamefont {S.}~\bibnamefont
  {Zhang}}, \bibinfo {author} {\bibfnamefont {A.}~\bibnamefont {Song}},
  \bibinfo {author} {\bibfnamefont {L.}~\bibnamefont {Chen}}, \bibinfo {author}
  {\bibfnamefont {C.}~\bibnamefont {Jiang}}, \bibinfo {author} {\bibfnamefont
  {C.}~\bibnamefont {Chen}}, \bibinfo {author} {\bibfnamefont {L.}~\bibnamefont
  {Gao}}, \bibinfo {author} {\bibfnamefont {Y.}~\bibnamefont {Hou}}, \bibinfo
  {author} {\bibfnamefont {L.}~\bibnamefont {Liu}}, \bibinfo {author}
  {\bibfnamefont {T.}~\bibnamefont {Ma}}, \bibinfo {author} {\bibfnamefont
  {H.}~\bibnamefont {Wang}}, \emph {et~al.},\ }\bibfield  {title} {\bibinfo
  {title} {Abnormal conductivity in low-angle twisted bilayer graphene},\
  }\href {https://doi.org/10.1126/sciadv.abc5555} {\bibfield  {journal}
  {\bibinfo  {journal} {Science advances}\ }\textbf {\bibinfo {volume} {6}},\
  \bibinfo {pages} {eabc5555} (\bibinfo {year} {2020})}\BibitemShut {NoStop}%
\bibitem [{\citenamefont {Leconte}\ \emph {et~al.}(2022)\citenamefont
  {Leconte}, \citenamefont {Javvaji}, \citenamefont {An}, \citenamefont
  {Samudrala},\ and\ \citenamefont {Jung}}]{tb1}%
  \BibitemOpen
  \bibfield  {author} {\bibinfo {author} {\bibfnamefont {N.}~\bibnamefont
  {Leconte}}, \bibinfo {author} {\bibfnamefont {S.}~\bibnamefont {Javvaji}},
  \bibinfo {author} {\bibfnamefont {J.}~\bibnamefont {An}}, \bibinfo {author}
  {\bibfnamefont {A.}~\bibnamefont {Samudrala}},\ and\ \bibinfo {author}
  {\bibfnamefont {J.}~\bibnamefont {Jung}},\ }\bibfield  {title} {\bibinfo
  {title} {Relaxation effects in twisted bilayer graphene: A multiscale
  approach},\ }\href
  {https://doi.org/https://doi.org/10.1103/PhysRevB.106.115410} {\bibfield
  {journal} {\bibinfo  {journal} {Physical Review B}\ }\textbf {\bibinfo
  {volume} {106}},\ \bibinfo {pages} {115410} (\bibinfo {year}
  {2022})}\BibitemShut {NoStop}%
\bibitem [{\citenamefont {Plimpton}(1995)}]{lammps1}%
  \BibitemOpen
  \bibfield  {author} {\bibinfo {author} {\bibfnamefont {S.}~\bibnamefont
  {Plimpton}},\ }\bibfield  {title} {\bibinfo {title} {Fast parallel algorithms
  for short-range molecular dynamics},\ }\href
  {https://doi.org/https://doi.org/10.1006/jcph.1995.1039} {\bibfield
  {journal} {\bibinfo  {journal} {Journal of computational physics}\ }\textbf
  {\bibinfo {volume} {117}},\ \bibinfo {pages} {1} (\bibinfo {year}
  {1995})}\BibitemShut {NoStop}%
\bibitem [{\citenamefont {Thompson}\ \emph {et~al.}(2022)\citenamefont
  {Thompson}, \citenamefont {Aktulga}, \citenamefont {Berger}, \citenamefont
  {Bolintineanu}, \citenamefont {Brown}, \citenamefont {Crozier}, \citenamefont
  {In't~Veld}, \citenamefont {Kohlmeyer}, \citenamefont {Moore}, \citenamefont
  {Nguyen} \emph {et~al.}}]{lammps2}%
  \BibitemOpen
  \bibfield  {author} {\bibinfo {author} {\bibfnamefont {A.~P.}\ \bibnamefont
  {Thompson}}, \bibinfo {author} {\bibfnamefont {H.~M.}\ \bibnamefont
  {Aktulga}}, \bibinfo {author} {\bibfnamefont {R.}~\bibnamefont {Berger}},
  \bibinfo {author} {\bibfnamefont {D.~S.}\ \bibnamefont {Bolintineanu}},
  \bibinfo {author} {\bibfnamefont {W.~M.}\ \bibnamefont {Brown}}, \bibinfo
  {author} {\bibfnamefont {P.~S.}\ \bibnamefont {Crozier}}, \bibinfo {author}
  {\bibfnamefont {P.~J.}\ \bibnamefont {In't~Veld}}, \bibinfo {author}
  {\bibfnamefont {A.}~\bibnamefont {Kohlmeyer}}, \bibinfo {author}
  {\bibfnamefont {S.~G.}\ \bibnamefont {Moore}}, \bibinfo {author}
  {\bibfnamefont {T.~D.}\ \bibnamefont {Nguyen}}, \emph {et~al.},\ }\bibfield
  {title} {\bibinfo {title} {Lammps-a flexible simulation tool for
  particle-based materials modeling at the atomic, meso, and continuum
  scales},\ }\href {https://doi.org/https://doi.org/10.1016/j.cpc.2021.108171}
  {\bibfield  {journal} {\bibinfo  {journal} {Computer Physics Communications}\
  }\textbf {\bibinfo {volume} {271}},\ \bibinfo {pages} {108171} (\bibinfo
  {year} {2022})}\BibitemShut {NoStop}%
\bibitem [{\citenamefont {Brenner}(1990)}]{rebo1}%
  \BibitemOpen
  \bibfield  {author} {\bibinfo {author} {\bibfnamefont {D.~W.}\ \bibnamefont
  {Brenner}},\ }\bibfield  {title} {\bibinfo {title} {Empirical potential for
  hydrocarbons for use in simulating the chemical vapor deposition of diamond
  films},\ }\href {https://doi.org/https://doi.org/10.1103/PhysRevB.42.9458}
  {\bibfield  {journal} {\bibinfo  {journal} {Physical review B}\ }\textbf
  {\bibinfo {volume} {42}},\ \bibinfo {pages} {9458} (\bibinfo {year}
  {1990})}\BibitemShut {NoStop}%
\bibitem [{\citenamefont {Brenner}\ \emph {et~al.}(2002)\citenamefont
  {Brenner}, \citenamefont {Shenderova}, \citenamefont {Harrison},
  \citenamefont {Stuart}, \citenamefont {Ni},\ and\ \citenamefont
  {Sinnott}}]{rebo2}%
  \BibitemOpen
  \bibfield  {author} {\bibinfo {author} {\bibfnamefont {D.~W.}\ \bibnamefont
  {Brenner}}, \bibinfo {author} {\bibfnamefont {O.~A.}\ \bibnamefont
  {Shenderova}}, \bibinfo {author} {\bibfnamefont {J.~A.}\ \bibnamefont
  {Harrison}}, \bibinfo {author} {\bibfnamefont {S.~J.}\ \bibnamefont
  {Stuart}}, \bibinfo {author} {\bibfnamefont {B.}~\bibnamefont {Ni}},\ and\
  \bibinfo {author} {\bibfnamefont {S.~B.}\ \bibnamefont {Sinnott}},\
  }\bibfield  {title} {\bibinfo {title} {A second-generation reactive empirical
  bond order (rebo) potential energy expression for hydrocarbons},\ }\href
  {https://doi.org/10.1088/0953-8984/14/4/312} {\bibfield  {journal} {\bibinfo
  {journal} {Journal of Physics: Condensed Matter}\ }\textbf {\bibinfo {volume}
  {14}},\ \bibinfo {pages} {783} (\bibinfo {year} {2002})}\BibitemShut
  {NoStop}%
\bibitem [{\citenamefont {Kolmogorov}\ and\ \citenamefont
  {Crespi}(2005)}]{kolmogorov}%
  \BibitemOpen
  \bibfield  {author} {\bibinfo {author} {\bibfnamefont {A.~N.}\ \bibnamefont
  {Kolmogorov}}\ and\ \bibinfo {author} {\bibfnamefont {V.~H.}\ \bibnamefont
  {Crespi}},\ }\bibfield  {title} {\bibinfo {title} {Registry-dependent
  interlayer potential for graphitic systems},\ }\href
  {https://doi.org/https://doi.org/10.1103/PhysRevB.71.235415} {\bibfield
  {journal} {\bibinfo  {journal} {Physical Review B}\ }\textbf {\bibinfo
  {volume} {71}},\ \bibinfo {pages} {235415} (\bibinfo {year}
  {2005})}\BibitemShut {NoStop}%
\bibitem [{\citenamefont {Fang}\ and\ \citenamefont {Kaxiras}(2016)}]{tb4}%
  \BibitemOpen
  \bibfield  {author} {\bibinfo {author} {\bibfnamefont {S.}~\bibnamefont
  {Fang}}\ and\ \bibinfo {author} {\bibfnamefont {E.}~\bibnamefont {Kaxiras}},\
  }\bibfield  {title} {\bibinfo {title} {Electronic structure theory of weakly
  interacting bilayers},\ }\href
  {https://doi.org/https://doi.org/10.1103/PhysRevB.93.235153} {\bibfield
  {journal} {\bibinfo  {journal} {Physical Review B}\ }\textbf {\bibinfo
  {volume} {93}},\ \bibinfo {pages} {235153} (\bibinfo {year}
  {2016})}\BibitemShut {NoStop}%
\bibitem [{\citenamefont {Kang}\ and\ \citenamefont {Vafek}(2018)}]{tb2}%
  \BibitemOpen
  \bibfield  {author} {\bibinfo {author} {\bibfnamefont {J.}~\bibnamefont
  {Kang}}\ and\ \bibinfo {author} {\bibfnamefont {O.}~\bibnamefont {Vafek}},\
  }\bibfield  {title} {\bibinfo {title} {Symmetry, maximally localized wannier
  states, and a low-energy model for twisted bilayer graphene narrow bands},\
  }\href {https://doi.org/https://doi.org/10.1103/PhysRevX.8.031088} {\bibfield
   {journal} {\bibinfo  {journal} {Physical Review X}\ }\textbf {\bibinfo
  {volume} {8}},\ \bibinfo {pages} {031088} (\bibinfo {year}
  {2018})}\BibitemShut {NoStop}%
\bibitem [{\citenamefont {Slater}\ and\ \citenamefont {Koster}(1954)}]{slater}%
  \BibitemOpen
  \bibfield  {author} {\bibinfo {author} {\bibfnamefont {J.~C.}\ \bibnamefont
  {Slater}}\ and\ \bibinfo {author} {\bibfnamefont {G.~F.}\ \bibnamefont
  {Koster}},\ }\bibfield  {title} {\bibinfo {title} {Simplified lcao method for
  the periodic potential problem},\ }\href
  {https://doi.org/https://doi.org/10.1103/PhysRev.94.1498} {\bibfield
  {journal} {\bibinfo  {journal} {Physical review}\ }\textbf {\bibinfo {volume}
  {94}},\ \bibinfo {pages} {1498} (\bibinfo {year} {1954})}\BibitemShut
  {NoStop}%
\bibitem [{\citenamefont {De~Laissardiere}\ \emph {et~al.}(2012)\citenamefont
  {De~Laissardiere}, \citenamefont {Mayou},\ and\ \citenamefont
  {Magaud}}]{param}%
  \BibitemOpen
  \bibfield  {author} {\bibinfo {author} {\bibfnamefont {G.~T.}\ \bibnamefont
  {De~Laissardiere}}, \bibinfo {author} {\bibfnamefont {D.}~\bibnamefont
  {Mayou}},\ and\ \bibinfo {author} {\bibfnamefont {L.}~\bibnamefont
  {Magaud}},\ }\bibfield  {title} {\bibinfo {title} {Numerical studies of
  confined states in rotated bilayers of graphene},\ }\href
  {https://doi.org/https://doi.org/10.1103/PhysRevB.86.125413} {\bibfield
  {journal} {\bibinfo  {journal} {Physical Review B}\ }\textbf {\bibinfo
  {volume} {86}},\ \bibinfo {pages} {125413} (\bibinfo {year}
  {2012})}\BibitemShut {NoStop}%
\bibitem [{\citenamefont {Datta}(1997)}]{negf1}%
  \BibitemOpen
  \bibfield  {author} {\bibinfo {author} {\bibfnamefont {S.}~\bibnamefont
  {Datta}},\ }\href@noop {} {\emph {\bibinfo {title} {Electronic transport in
  mesoscopic systems}}}\ (\bibinfo  {publisher} {Cambridge university press},\
  \bibinfo {year} {1997})\BibitemShut {NoStop}%
\bibitem [{\citenamefont {Datta}(2005)}]{negf2}%
  \BibitemOpen
  \bibfield  {author} {\bibinfo {author} {\bibfnamefont {S.}~\bibnamefont
  {Datta}},\ }\href@noop {} {\emph {\bibinfo {title} {Quantum transport: atom
  to transistor}}}\ (\bibinfo  {publisher} {Cambridge university press},\
  \bibinfo {year} {2005})\BibitemShut {NoStop}%
\bibitem [{\citenamefont {Di~Ventra}(2008)}]{negf3}%
  \BibitemOpen
  \bibfield  {author} {\bibinfo {author} {\bibfnamefont {M.}~\bibnamefont
  {Di~Ventra}},\ }\href@noop {} {\emph {\bibinfo {title} {Electrical transport
  in nanoscale systems}}}\ (\bibinfo  {publisher} {Cambridge University
  Press},\ \bibinfo {year} {2008})\BibitemShut {NoStop}%
\bibitem [{\citenamefont {Cresti}\ \emph {et~al.}(2003)\citenamefont {Cresti},
  \citenamefont {Farchioni}, \citenamefont {Grosso},\ and\ \citenamefont
  {Parravicini}}]{curr1}%
  \BibitemOpen
  \bibfield  {author} {\bibinfo {author} {\bibfnamefont {A.}~\bibnamefont
  {Cresti}}, \bibinfo {author} {\bibfnamefont {R.}~\bibnamefont {Farchioni}},
  \bibinfo {author} {\bibfnamefont {G.}~\bibnamefont {Grosso}},\ and\ \bibinfo
  {author} {\bibfnamefont {G.~P.}\ \bibnamefont {Parravicini}},\ }\bibfield
  {title} {\bibinfo {title} {Keldysh-green function formalism for current
  profiles in mesoscopic systems},\ }\href
  {https://doi.org/https://doi.org/10.1103/PhysRevB.68.075306} {\bibfield
  {journal} {\bibinfo  {journal} {Physical Review B}\ }\textbf {\bibinfo
  {volume} {68}},\ \bibinfo {pages} {075306} (\bibinfo {year}
  {2003})}\BibitemShut {NoStop}%
\bibitem [{\citenamefont {Caroli}\ \emph {et~al.}(1971)\citenamefont {Caroli},
  \citenamefont {Combescot}, \citenamefont {Nozieres},\ and\ \citenamefont
  {Saint-James}}]{curr2}%
  \BibitemOpen
  \bibfield  {author} {\bibinfo {author} {\bibfnamefont {C.}~\bibnamefont
  {Caroli}}, \bibinfo {author} {\bibfnamefont {R.}~\bibnamefont {Combescot}},
  \bibinfo {author} {\bibfnamefont {P.}~\bibnamefont {Nozieres}},\ and\
  \bibinfo {author} {\bibfnamefont {D.}~\bibnamefont {Saint-James}},\
  }\bibfield  {title} {\bibinfo {title} {Direct calculation of the tunneling
  current},\ }\href {https://doi.org/10.1088/0022-3719/4/8/018} {\bibfield
  {journal} {\bibinfo  {journal} {Journal of Physics C: Solid State Physics}\
  }\textbf {\bibinfo {volume} {4}},\ \bibinfo {pages} {916} (\bibinfo {year}
  {1971})}\BibitemShut {NoStop}%
\bibitem [{\citenamefont {Edwards}\ and\ \citenamefont {Thouless}(1972)}]{ipr}%
  \BibitemOpen
  \bibfield  {author} {\bibinfo {author} {\bibfnamefont {J.}~\bibnamefont
  {Edwards}}\ and\ \bibinfo {author} {\bibfnamefont {D.}~\bibnamefont
  {Thouless}},\ }\bibfield  {title} {\bibinfo {title} {Numerical studies of
  localization in disordered systems},\ }\href
  {https://doi.org/10.1088/0022-3719/5/8/007} {\bibfield  {journal} {\bibinfo
  {journal} {Journal of Physics C: Solid State Physics}\ }\textbf {\bibinfo
  {volume} {5}},\ \bibinfo {pages} {807} (\bibinfo {year} {1972})}\BibitemShut
  {NoStop}%
\end{thebibliography}
\end{document}